\documentclass[10pt, journal,compsoc]{IEEEtran}

\usepackage{cite}
\usepackage{verbatim}
\usepackage{amsmath}
\usepackage{algorithmic}
\usepackage{array}
\usepackage[caption=false,font=footnotesize]{subfig}
\usepackage{url}
\usepackage{graphicx}
\usepackage{amsthm, amssymb}
\usepackage{textcomp}
\usepackage{xcolor}
\usepackage[utf8]{inputenc}
\usepackage[T1]{fontenc}
\usepackage{balance}
\usepackage{xspace}
\usepackage{multirow}
\usepackage{float}
\usepackage{soul}
\usepackage{paralist}
\usepackage{url}
\usepackage{hyperref}

\newboolean{showcomments}
\setboolean{showcomments}{true}         
\ifthenelse{\boolean{showcomments}}
{
}
{
	
}

\theoremstyle{definition}

\begin{document}
\title{\textsc{Prevent}\xspace: An Unsupervised Approach to Predict Software Failures in Production}

\author{\IEEEauthorblockN{Giovanni Denaro$^*$, Rahim  Heydarov$^{+}$, Ali Mohebbi$^{+}$, Mauro Pezz\`e$^{+,\#}$}
\IEEEauthorblockA{
\begin{tabular}{ccc}
~\\
$^{*}$ University of Milano-Bicocca & $^{+}$ Universit\`a della Svizzera Italiana (USI)& $^{\#}$ Constructor Institute\\
Milano, Italy & Lugano, Switzerland & Schaffhausen, Switzerland\\
~\\
\end{tabular}\\
Email:  giovanni.denaro@unimib.it, \{rahim.heydarov, ali.mohebbi, mauro.pezze\}@usi.ch}
}

\IEEEtitleabstractindextext{\begin{abstract}
This paper presents  \textsc{Prevent}\xspace, a fully unsupervised approach to predict and localize failures in distributed enterprise applications.

Software failures in production are unavoidable. 
Predicting failures and locating failing components online are the first steps to proactively manage faults in production.
Many techniques predict failures from anomalous combinations of system metrics with supervised, weakly supervised, and semi-supervised learning models.
Supervised approaches require large sets of labelled data not commonly available in large enterprise applications, and address failure types that can be either captured with predefined rules or observed while training supervised models.  

\textsc{Prevent}\xspace integrates the core ingredients of unsupervised approaches into a novel fully unsupervised approach to  predict failures and localize failing resources.  
The results of experimenting with \mbox{\textsc{Prevent}\xspace} on a commercially-compliant distributed cloud system indicate that \mbox{\textsc{Prevent}\xspace} provides more stable, reliable and timely predictions than supervised learning approaches, without requiring the often impractical training with labeled data.

\end{abstract}
}

\maketitle

\section{Introduction}\label{sec:intro}

Good design and quality assurance practice cannot prevent software to fail in production~\cite{Gazzola:field:ISSRE:2017}. The complexity of distributed enterprise applications further increases the risks of production failures.

\textcolor{black}{Several approaches apply machine learning techniques to prevent system failures, following many studies that exploit artificial intelligence for software engineering problems~\cite{tu:frugal:ase:2021,zhang:lowrank:iet:2021,lu:defect:ase:2012,
Zhang:WeakSupervision:CoRR2022,Bach:Learning:GenerativeModelsWithoutLabeledData:ICML:2017,Liang:LearningFromMeasurements:ICML:2009,Ratner:LargeTrainingSetsQuickly:ANIPS:2016,Pan:TransferLearningSurvey:IEEE:2010,Stewart:LabelFreeSupervision:AAAI:2017,
Druck:ActiveLearning:EMNLP:2009,
Chapelle:SemiSupervisedLearning:MIT:2006,Berthelot:MixmatchSemiSupervisedLearning:ANIPS:2019,Laine:TemporalEnsemblingSemiSupervisedLearning:arXiv:2016,Mann:SemiSupervisedLearningWeaklyLabeled:MLR:2010,
bodo:process:serp:2015, 
denaro:fault-proneness:seke:2002}.}
\textcolor{black}{The two mainstream classes of approaches either predict failure prone modules based on metrics that reflect the complexity of the code or predict the occurrence of error states at runtime based on metrics that reflect the execution of software systems.}

\textcolor{black}{Approaches that predict failure prone modules feed prediction models with metrics  from the code, and  allow for fine-tuning testing activities for the modules that are more likely to contain defects~\cite{denaro:fault-proneness:seke:2002,zhang:lowrank:iet:2021}.
A notable case of approaches that predict failure prone modules is the Nam and Kim's CLAMI approach~\cite{Nam:UnlabeledDefectPrediction:IEEE:2015} that clusters software modules based on the similarity of code complexity metrics.}

\textcolor{black}{Approaches that predict the occurrence of error states at runtime feed prediction models with metrics  monitored at runtime, and allow to take countermeasures before the failures actually manifest.} These approaches draw on the observation that
many failures occur in production when the execution of some faulty statements corrupts the execution state, and eventually the error state propagates to a system failure, that is, a deviation of the delivered service from the required functionality.

Current approaches for predicting failures exploit rule-based, signature-based, or semi-supervised strategies.
\emph{Rule-based approaches} rely on predicates that experts extract from data 
observed during operations~\cite{Chung:bottleneckdetection:IPDPS:2008}.
\emph{Signature-based approaches} rely on supervised learning models that leverage the information from historical records of previously observed  failures~\cite{Bodik:fingerprinting:EUROSYS:2010,Malik:AutomaticDetection:ICSE:2013,nistor:suncat:issta:2014,ozcelik:seer:tse:2016,Sauvanaud:FaultLocalizationClearwater:ISSRE:2016,tehrani:threshold-sensitive:jmgs:2017,islam:failure-in-cloud:iccc:2017,davis:failuresim:cloud:2017,gao:task-failure:bigdata:2019,sun:hardware-failure:dac:2019}.
\textcolor{black}{Signature-based approaches require large amounts of labeled failure data for training, which are rarely available and hard to collect.}
\emph{Semi-supervised approaches} exploit signature-based models on top of synthetic data inferred with either semi-supervised, \textcolor{black}{weakly supervised} or unsupervised learning, to balance accuracy and required information~\cite{Fulp:PredictingFailure:WASL:2008,tan:prepare:icdcs:2012,Tan:AnomalyPrediction:PODC:2010,guan:ensemble:jcom:2012,Mariani:PreMiSE:JSS:2020}.
Signature-based approaches for distributed applications often aim also to localize the components responsible for the  failures~\cite{Chung:bottleneckdetection:IPDPS:2008,Tan:AnomalyPrediction:PODC:2010,Magalhaes:rootcause:SAC:2011,Sambasivan:diagnosingperformance:NSDI:2011,Ibidunmoye:AnomalyDetectionSurvey:2015,Mariani:PreMiSE:JSS:2020,Mariani:LOUD:ICST:2018}.

\textcolor{black}{This paper investigates a \emph{purely unsupervised} approach to failure prediction.}
 Whereas signature-based approaches  
identify failures of pre-defined failure types that occur in the training data,
\emph{unsupervised} approaches detect anomalies as deviations from some model of the nominal system behavior suitably inferred in absence of failures, and can thus reveal failures of any types, including types that do not necessarily correspond to training data. 
\textcolor{black}{Previous work explored unsupervised anomaly detection for intrusion attacks~\cite{Bontemps:anomalydetection:FDSE:2016}, anomalies of streaming data~\cite{Ahmad:HTM:Neurocomputing:2017}, issues in computer networks~\cite{Fernandes:anomalydetection:JNCA:2016}, log file analysis~\cite{Du:deeplog:CCS:2017} and detection of performance issues~\cite{Ibidunmoye:AnomalyDetection:TNSM:2018}.}

\textcolor{black}{In this paper we propose \textsc{Prevent}\xspace, an original application of unsupervised approaches to predict and localize failures in distributed enterprise applications.}
\textcolor{black}{\textsc{Prevent}\xspace is grounded on the lessons learned from  \textsc{PreMiSe}\xspace, \textsc{EmBeD}\xspace and \textsc{Loud}\xspace, our previous work on supervised and unsupervised approaches to predict failures and localize faults.}

\textcolor{black}{
\textsc{PreMiSe}\xspace~\cite{Mariani:PreMiSE:JSS:2020} combines anomaly detection technique and Logistic Model Trees (LMT) to predict failures and localize failing resources in supervised fashion.
\textsc{EmBeD}\xspace~\cite{monni:energy:NIER:2019,monni:RBM:ICST:2019} proposes Restricted Boltzmann Machine (RBM) to predict failures in complex cloud systems.
\textsc{Loud}\xspace~\cite{Mariani:LOUD:ICST:2018} combines Granger causality analysis and page rank centrality measures to localize failing components. 
While \textsc{PreMiSe}\xspace offers a supervised end-to-end solution to predict failures and localize faults, \textsc{EmBeD}\xspace and \textsc{Loud}\xspace propose unsupervised approaches to predict failures and locate faults, respectively.}
\textcolor{black}{Predicting the occurrence of failures at the system level without localizing the failing components in distributed applications (as in \textsc{EmBeD}\xspace) does not offer developers and maintainers  enough information to activate prevention mechanisms.
Relying only on localization mechanisms (as in \textsc{Loud}\xspace) results in many false alarms.}

\textcolor{black}{\textsc{Prevent}\xspace originally 
combines failure prediction and failure localization in a coordinated  and fully unsupervised approach.
It exploits 
a deep autoencoder model to predict failures, and coordinates deep autoencoder, Granger causality and page rank centrality to locate failing components.  } 
\textcolor{black}{In a nutshell, 
The deep autoencoder identifies anomalous states of the system, and discriminates normal and anomalous metrics. 
The Granger causality analysis combined with page rank centrality effectively ranks the values that the deep autoencoder reveals as anomalous, to spot the faulty components responsible for the predicted failures.}

We introduce a new large dataset that we obtained by running many experiments on \textsc{Redis}\xspace, a commercially-compliant, distributed cloud system.
The experiments that we discuss in the paper compare the stability, reliability and earliness of \textsc{Prevent}\xspace with \textsc{PreMiSe}\xspace, \textcolor{black}{\textsc{EmBeD}\xspace and \textsc{Loud}\xspace}, in the context of failures that we seed in \textsc{Redis}\xspace.

\textcolor{black}{\textsc{Prevent}\xspace recomputes the prediction at each time interval.}
We measure the stability of a prediction as the true positive rate after the first true prediction, that is, the continuity in correctly reporting the failing component in the presence of error states.  
\textcolor{black}{Intuitively, predictions consecutively raised at all timestamps over a time interval are clearer messages of failures than intermittent predictions that appear and disappear in a time interval}.
We measure the reliability of the prediction with reference to the false positive rate, that is, the frequency of alarms that do not correspond to real error states or wrongly track the failures to non-failing components: The lower the false positive rate, the higher the reliability of the prediction. 
\textcolor{black}{Intuitively, predictions are more effective when they occur only in error states than when they occur both in error and correct states.}
We measure the earliness of a prediction as the time interval between the first correct prediction and the actual failure, that is, the time interval for a proactive action on the failing component to prevent a failure before its occurrence.

This paper contributes to the research in software engineering by

\begin{itemize}
\item \textcolor{black}{defining \textsc{Prevent}\xspace, a new unsupervised approach that originally combines 
deep autoencoder, Granger causality and page rank centrality to predict failures and locate the corresponding faulty components.} 

\item presenting a large set of data collected from a commercially-compliant, distributed cloud systems to evaluate and compare different approaches, data that we offer in a replication package,\footnote{The replication package at~\href{https://star.inf.usi.ch/\#/software-data/14}{https://star.inf.usi.ch/\#/software-data/14}\@\xspace}
\item comparing \textsc{Prevent}\xspace with \textsc{PreMiSe}\xspace, \textcolor{black}{\textsc{EmBeD}\xspace and \textsc{Loud}\xspace,}
to indicate the advantages of a fully unsupervised approach for both predicting failures and locating faulty components with respect to supervised approaches.
\end{itemize}

This paper is organized as follows. 
Section~\ref{sec:approach} presents \textsc{Prevent}\xspace. 
Section~\ref{sec:experiments} describes the experimental setting, and discusses the results of the experiments that comparatively evaluate \textsc{Prevent}\xspace with respect to \textsc{PreMiSe}\xspace, \textsc{EmBeD}\xspace and \textsc{Loud}\xspace.
Section~\ref{sec:related} discusses the main state-of-the-art approaches for predicting and diagnosing failures, and their relation with \textsc{Prevent}\xspace.
Section~\ref{sec:conclusions} summarizes the contribution of the paper and indicates novel research directions.

\section{\textsc{Prevent}\xspace}\label{sec:approach}

\textcolor{black}{The core contribution of this paper is \textsc{Prevent}\xspace, an unsupervised approach that both predicts failures in distributed enterprise applications and localizes the corresponding faulty components.  \textsc{Prevent}\xspace originally combines 
a deep autoencoder, Granger causality analysis and pagerank centrality analysis to predict failures and localize faulty components without requiring training with labeled data. With this original combination of unsupervised techniques, \textsc{Prevent}\xspace overcomes the main obstacles of supervised approaches: 
the effort required to label the data for training 
\textcolor{black}{and the difficulty of predicting failures of types that do not correspond to training data}. The unsupervised nature of \textsc{Prevent}\xspace allows to both predict failures of any type and locate the faulty components responsible for the failures.
Efficiently locating the faulty components in large distributed enterprise applications provides enough information for state-of-the-art self healing approaches~\cite{stack:selfhealing:CloudNG:2017} to automatically activate healing actions before the occurrence of  the failures.}

Figure~\ref{fig:approach:overall} shows the main components of \textsc{Prevent}\xspace, the \emph{State classifier} and the \emph{Anomaly ranker}, that predict failures and localize the faulty components, respectively. 
As shown in the figure, the \emph{State classifier} and the \emph{Anomaly ranker} work on time series of Key Performance Indicators, \emph{KPI}s, that are sets of metric values observed by monitoring the application at regular time intervals (every minute in our experiments).
A \emph{KPI} is a pair $\langle metric,node\rangle$ of a metric value collected at either a virtual or physical node of the monitored application.  
Our current prototype of
\textsc{Prevent}\xspace collects \emph{KPI} series with a monitoring facility built on top of  Elasticsearch~\cite{elasticsearch:elastic.co}.

\begin{figure}[t!]
\begin{center}
\includegraphics[width=88mm]{imgs/Approach_new}\\
\caption{Overview of \textsc{Prevent}\xspace}
\label{fig:approach:overall}
\end{center}
\end{figure}

\textsc{Prevent}\xspace returns a list of anomalous nodes ranked by anomaly relevance, list that the \emph{Anomaly ranker} produces in the presence of anomalous states inferred with the \emph{State classifier}.

\textcolor{black}{Both the \emph{State classifier} and the \emph{Anomaly ranker} include 
a deep autoencoder that requires unsupervised training with KPI data collected during normal (non-failing) executions of the application, without requiring any labels at training time.  
The deep autoencoder can be trained with a reasonably small amount of data. 
In our experiments we train \textsc{Prevent}\xspace with data collected in two weeks of normal execution. Thus, \textsc{Prevent}\xspace is resilient to concept drifts that span over several weeks or more.  Such drifts can be overtaken with both frequent short retraining sessions and continuous training in production, thanks to the unsupervised nature of \textsc{Prevent}\xspace.}

The \emph{State classifier} crosschecks the KPI values observed in production against the normal-execution characteristics inferred during training, and accepts \emph{normal states}, when there are no significant differences.  It pinpoints \emph{anomalous states}, otherwise.
The \emph{Anomaly ranker} identifies anomalous KPIs, that is, KPIs with values that significantly differ from values observed during training in normal execution conditions, and ranks anomalous nodes according to their relevance with respect to the anomalous KPIs.  \textsc{Prevent}\xspace reports the anomalous nodes only when the  \emph{State classifier} reveals anomalous states.
We discuss the inference models that instantiate the two components later in this section.

\textcolor{black}{We defined \textsc{Prevent}\xspace by benefitting from the lessons learned with \textsc{PreMiSe}\xspace~\cite{Mariani:PreMiSE:JSS:2020},  \textsc{EmBeD}\xspace~\cite{monni:RBM:ICST:2019, monni:energy:NIER:2019}, and \textsc{Loud}\xspace~\cite{Mariani:LOUD:ICST:2018}, three representative techniques to predict and localize failures.}

\label{line:premise} \textcolor{black}{The supervised \textsc{PreMiSe}\xspace approach that we developed as a joint project with industrial partners, gives us important insights about the strong limitations of supervised approaches in many industrially-relevant domains.  
\textsc{PreMiSe}\xspace combines an unsupervised approach for detecting anomalous KPIs with a supervised signature-based approach for predicting failures.  The signature-based approach requires long supervised training, and achieves good precision only for failures of types considered during supervised training.
\textsc{PreMiSe}\xspace indicates that supervised approaches can indeed precisely and timely predict failures, localize faults and identify the fault types. It also highlights the strong limitations of training systems with seeded faults in production, as supervised approaches require.     
}

\textcolor{black}{\textsc{EmBeD}\xspace predicts failures by both computing the Gibbs free energy associated with the KPIs that represent the system state and monitoring anomalous energy fluctuations in production.
\textsc{EmBeD}\xspace provides evidence about the correlation between anomalous energy values of the RBM and KPI anomalies that can lead to system failures.
\textsc{EmBeD}\xspace predicts failures of the target application as a whole, without any information at the level of the application nodes.}

\textcolor{black}{\textsc{Loud}\xspace localizes faults with an unsupervised observational strategy,  by identifying the application nodes that are highly relevant with respect to the causal dependencies between the observed anomalies. 
It assumes the availability of precise  anomaly predictions at system-level to limit the otherwise large amount of false alarms. 
\textsc{Loud}\xspace shows how to combine Granger causality analysis and page rank centrality to effectively locate faulty components, in the presence of reliable failure predictions.}

\textcolor{black}{\textsc{Prevent}\xspace introduces a deep autoencoder to improve the precision of detecting and classifying anomalous KPIs, and proposes an original combination of the approaches as an efficient end-to-end solution to predict failures and locate faulty components with lightweight unsupervised training.}

\textcolor{black}{We implemented two \emph{State classifier}s that we refer to as \textsc{Prevent$_A$}\xspace ($Prevent_{Autoencoder}$)  and \textsc{Prevent$_E$}\xspace ($Prevent_{Energy}$).  \textsc{Prevent$_A$}\xspace implements \textsc{Prevent}\xspace as illustrated in Figure~\ref{fig:approach:overall}.  
\textsc{Prevent$_E$}\xspace replaces the Deep Autoencoder of the \emph{State classifier} in Figure~\ref{fig:approach:overall} with a Restricted Boltzmann Machine (RBM) that implements the free-energy-based approach of \textsc{EmBeD}\xspace.  We use \textsc{Prevent$_E$}\xspace to compare \textsc{Prevent}\xspace with \textsc{EmBeD}\xspace.} 
Both \textsc{Prevent$_A$}\xspace and \textsc{Prevent$_E$}\xspace are integrated with the \emph{Anomaly ranker} that combines Granger-causality with eigenvector-centrality (page rank centrality calculator) to precisely localize failing nodes.

\subsection{\textsc{Prevent$_A$}\xspace State Classifier}

The \textsc{Prevent$_A$}\xspace State classifier uses a deep autoencoder to predict failures without requiring training with seeded faults.
The deep autoencoder model identifies anomalous KPIs as KPIs with values that are anomalous with respect to the observations when training with normal executions.

A deep autoencoder (also simply referred to as autoencoder) is a neural network architected with two contiguous sequences of layers that mirror each other structure:
A first sequence of layers of decreasing size up to an intermediate layer of minimal size, and a second sequence of layers of correspondingly increasing size up to a layer with the same size of the initial layer.

During training, the first half of the network learns how to encode the input data in incrementally condensed form, up to a minimal form in the intermediate layer. The second half of the network learns how to regenerate the input based on the information condensed in the intermediate layer. 
The difference between the input and output values is the \emph{reconstruction error} of the autoencoder. 

During training the neurons of the network learn functions that minimize the  average reconstruction error on the training data.
In production, the network returns small reconstruction errors for data similar to the training data in both  absolute values and mutual correlations. It returns large reconstruction errors for data that significantly differ from the observations in the training phase.

\begin{figure}[]
\begin{center}
\includegraphics[width=80mm]{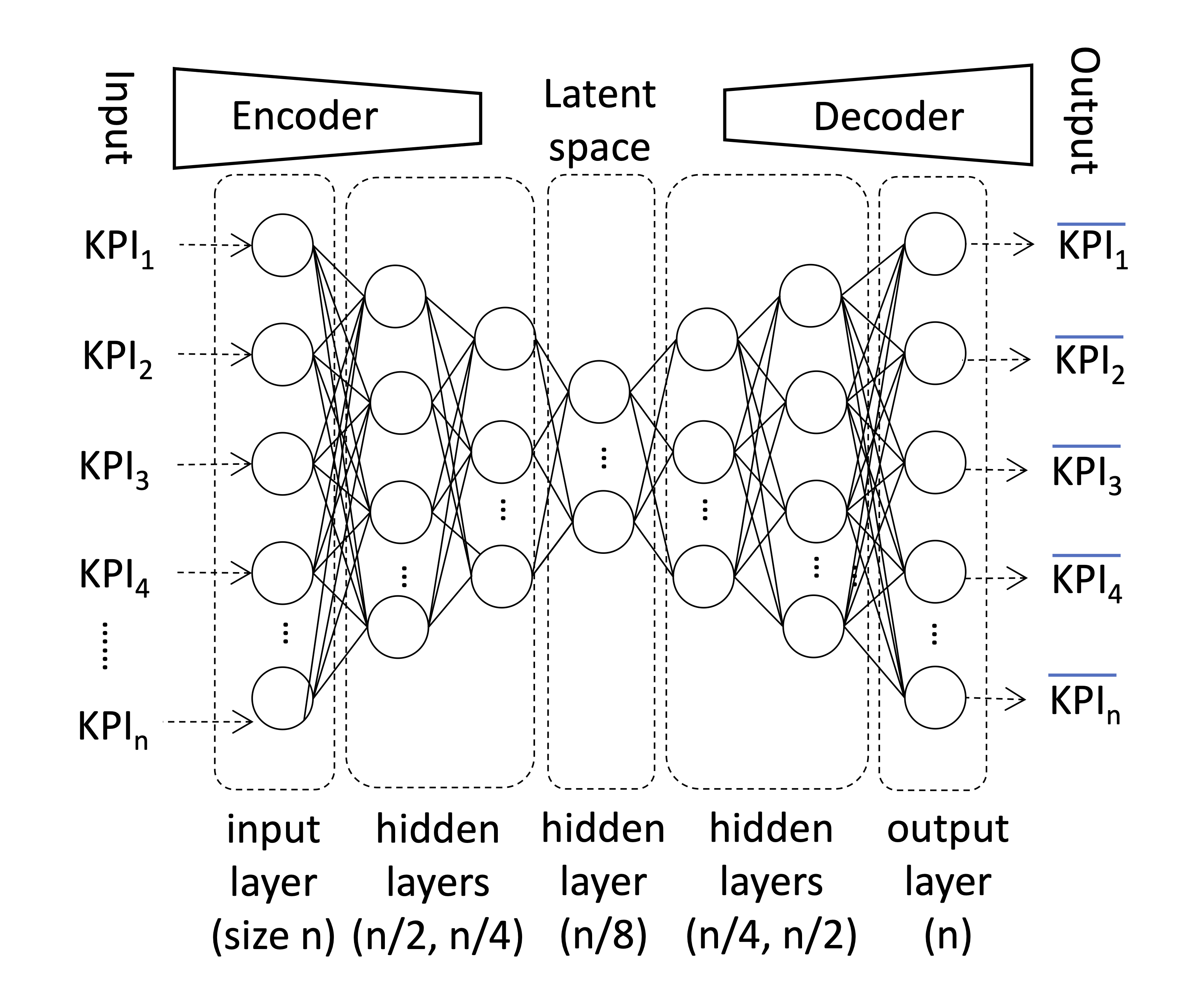}
\caption{Deep autoencoder as instantiated in \textsc{Prevent$_A$}\xspace} 
\label{fig:autoencoder}
\end{center}
\end{figure}

Figure~\ref{fig:autoencoder} illustrates the architecture of the \textsc{Prevent$_A$}\xspace autoencoder that is
composed of seven layers with sizes $n$, $n/2$,  $n/4$,  $n/8$, $n/4$, $n/2$, and $n$, respectively, being $n$ is the number of monitored KPIs.

We trained the \textsc{Prevent$_A$}\xspace autoencoder with the KPIs observed at regular time intervals on the distributed enterprise application executed in normal conditions, that is, without failures. 
The trained autoencoder reveals the anomalous states that emerge in production, as the states that correspond to
reconstruction errors that significantly differ from the mean reconstruction error computed during training. \textcolor{black}{In our current prototype this threshold is set to the reconstruction errors that 
differ from the mean reconstruction error for more than three times the standard deviation observed on the training set.}

\subsection{\textsc{Prevent$_E$}\xspace State Classifier}
\label{sec:state-classifier}

The \textsc{Prevent$_E$}\xspace State classifier infers anomalous combinations of KPI values from perturbations of the Gibbs free energy computed on time series of KPI values monitored from production. 

The Gibbs free energy was originally introduced in statistical physics to model macroscopic many-particle systems as a statistical characterization of the properties of single particles that affect the global physical quantities such as energy or temperature~\cite{Chandler:StatisticalMechanics:1987}, and is applied in many  domains, such as the growth of the World Wide Web and the spread of epidemics~\cite{Dorogovtsev:CriticalPhenomenaNetworks:RevModPhys:2008}.

We rely on the intuition that complex distributed enterprise software applications and physical systems share the dependencies of  properties of interest of the global state of the system (energy and temperature in the case of physical systems, failures in the case of software applications) on the collective configuration of basic elements (particles in the case of physical systems, KPI values in the case of software applications).  
\textcolor{black}{Intuitively, the execution of some faulty code produces some error states with anomalous KPI values that propagate through the execution, thus leading to a progressive alignment of anomalous KPI values.}   
Following this intuition, we recast the problem of predicting failures in complex distributed enterprise software applications to the problem of revealing the collective alignment of the KPIs of an application to correlated anomalous values. 

\textsc{Prevent$_E$}\xspace supersedes the intractability of analytically representing the physical dependencies that concretely govern the correlations among KPIs~\cite{Carreira-Perpinan:ContrastiveDivergence:AISTATS:2005} by approximating the computation of the Gibbs energy with restricted Bolzmann machines, \emph{RBM}~\cite{fischer:rbm-intro:springer:2012}, and signals anomalous states when the energy exceeds a threshold.
   
\begin{figure}[]
\begin{center}
\includegraphics[width=65mm]{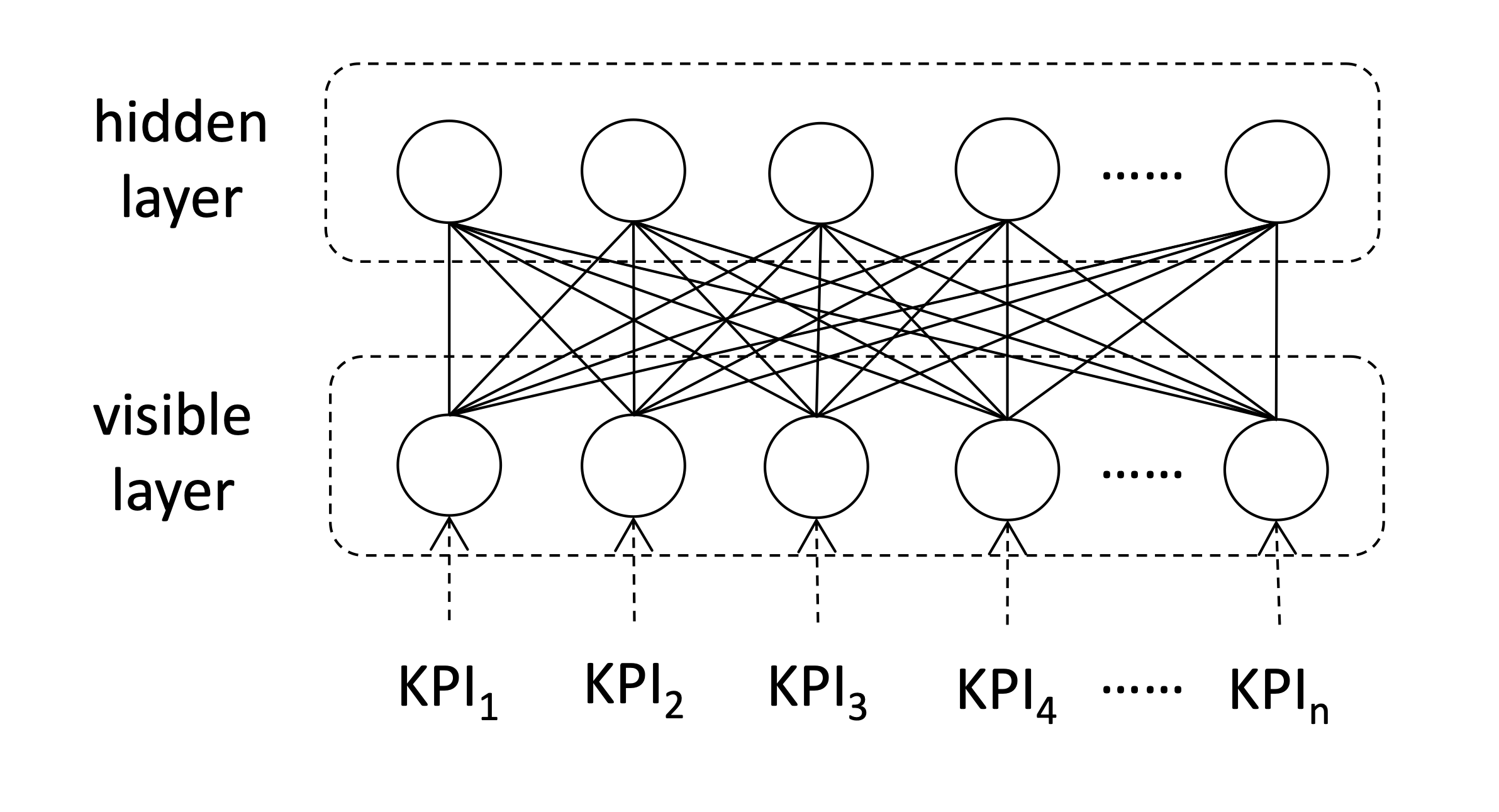}
\caption{RBM as instantiated in \textsc{Prevent$_E$}\xspace} 
\label{fig:rbm}
\end{center}
\end{figure}

Figure~\ref{fig:rbm} show the two convolutional layers of the RBM as instantiated in \textsc{Prevent$_E$}\xspace: A visible layer with as many neurons as the number of KPIs monitored on the target application, and a hidden layer with an equivalent number of neurons.
The visible layer takes in input the values of the KPIs monitored at each timestamp, while the hidden layer encodes the joint distributions of KPIs, based on the training-phase sampling of the conditional probabilities of the hidden nodes given the visible nodes. 

\textcolor{black}{\textsc{Prevent$_E$}\xspace trains the RBM with KPI values collected at regular time intervals in production to set the reference energy value.  Our experiments indicate that a training with KPI values collected over two weeks produces good results. \textsc{Prevent$_E$}\xspace reports anomalies when the energy value observed at runtime exceed the reference energy value by some thresholds. The threshold values can be computed offline by training the RBM with datasets from different applications.}
\textsc{Prevent$_E$}\xspace implements the RBM neural network in Matlab.

\subsection{\textsc{Prevent}\xspace Anomaly ranker}

The \textsc{Prevent}\xspace Anomaly ranker:

\begin{enumerate}[(i)]

\item  identifies sets of anomalous KPIs with a \emph{deep autoencoder}, 

\item builds a graph that represents the causality dependencies between the anomalous KPIs with \emph{Granger-causality analysis}~\cite{Profillidis:ModelingTransportDemand:Elsevier:2018, Arnold:TCMGranger:KDD:2007, Granger:causality:Econometrica:1969}, 

\item exploits the causality graph to calculate the \emph{PageRank centrality} value of each anomalous KPI in the graph~\cite{Langville:SurveyEigenvectorMethods:SIAM:2005, Scott:SNAnalysis:book:2011, Martin:Localization:PhRvE:2014},

\item returns the \emph{three nodes of the application with the highest numbers of anomalous KPIs}, selecting the top anomalous KPIs of the centrality ranking, up considering at most 20\% of the KPIs.

\end{enumerate}

\subsubsection*{Deep Autoencoder}

The  deep autoencoder of \textsc{Prevent}\xspace Anomaly ranker is the same deep autoencoder of \textsc{Prevent$_A$}\xspace state classifier.
\textcolor{black}{It identifies anomalous KPIs as KPIs with locally high reconstruction errors that we implement as KPI reconstruction errors that differ from the corresponding mean observed on the training set for more than three times the corresponding  standard deviation, in our prototype implementation.}

\subsubsection*{Granger Causality Analysis}

At each timestamp in production, the Granger-causality analyzer 
builds a causality graph that represents the causal dependencies among the KPIs with anomalous values at the considered timestamp. 
During training, \textsc{Prevent}\xspace builds a \emph{baseline causality graph} that represents the  causality relations between  the KPIs, as captured under normal execution conditions:
For each pair of KPIs $\langle k_a, k_b\rangle$, there is an edge from from the corresponding node $k_a$ to node $k_b$ in the baseline casualty graph, if the analysis of the time series of the two KPIs reveals a causal dependency from the values of $k_a$ on the values of $k_b$, according to the Granger causality test~\cite{Granger:causality:Econometrica:1969}. The weight of the edges indicates the strength of the causal dependencies.\footnote{
The Granger causality test determines the existence of  a causal dependency between a pair of KPIs $k_a$ and $k_b$ as a statistical test of whether the time series values of $k_a$ provide statistically significant information about the evolution of the future values of $k_b$~\cite{Profillidis:ModelingTransportDemand:Elsevier:2018, Arnold:TCMGranger:KDD:2007, Granger:causality:Econometrica:1969}. 
Specifically, we test the null-hypothesis that $k_a$ does not Granger-cause $k_b$ by
\begin{inparaenum}[(i)]
\item building an auto-regression model for the time series values of $k_b$,
\item building another regression model that considers the values of both $k_a$ and $k_b$ as predictors for the future values of $k_b$, 
\item testing if the latter model provides significantly better predictions than the former one.
\end{inparaenum}
If so, we reject the null-hypothesis in favor of the alternative hypothesis that $k_a$ Granger-causes $k_b$,
and compute the strength of the causality relation as the coefficient of determination $R^2$.
We implemented the test with the Statsmodels Python library~\cite{Seabold:statsmodels:SciPy:2010, McKinney:statsmodels:Jarrodmillman:2011}.}
At each timestamp in production, \textsc{Prevent}\xspace Anomaly ranker derives the causality graph of the anomalous KPIs by pruning the baseline causality graph, to exclude the KPIs that autoencoder does not indicate as anomalous. 

\subsubsection*{PageRank Centrality}

At each timestamp in production, the \textsc{Prevent}\xspace Anomaly ranker exploits
the causality graph of the anomalous KPIs 
to weight the relative relevance of the anomalous KPIs as the PageRank centrality index.
PageRank scores the graph nodes (KPIs) according to both the number of incoming edges and the probability of anomalies to randomly spread through the graph (teleportation)~\cite{Langville:SurveyEigenvectorMethods:SIAM:2005}.

\subsubsection*{Top Anomalous Application Nodes}

\textsc{Prevent}\xspace Anomaly ranker sorts the anomalous KPIs according to the decreasing values of their centrality scores; It selects the top anomalous KPIs of the ranking up to considering at most 20\% of the KPIs; It tracks these anomalous KPIs to the corresponding nodes of the application, and returns the three application nodes with highest number of top-anomalous KPIs that correspond to their location.

\smallskip

The \textsc{Prevent}\xspace Anomaly ranker improves \textsc{Loud}\xspace by 
\begin{inparaenum}[(i)]
\item discriminating the anomalous KPIs with a deep autoencoder instead of time-series analysis, to improve the results,
\item exploiting the top anomalous KPIs of the ranking up to 20\% of KPIs, rather than simply referring to the top 20 anomalous KPIs, to scale to large systems with thousands KPIs and
\item considering the KPIs that are anomalous at each specific timestamp, rather than the KPIs that are detected as anomalous at least once in the scope of a time window, to identify anomalous states and anomalous nodes at each timestamp.
\end{inparaenum}

\section{Experiments}\label{sec:experiments}

\subsection{Research Questions}
Our experiments address three research questions:

\begin{enumerate}[\textit{RQ}1:]
\item Can \textsc{Prevent}\xspace predict failures in distributed enterprise applications? 
\item Does the unsupervised \textsc{Prevent}\xspace approach improve over state-of-the-art (supervised) approaches?

\item Does \textsc{Prevent}\xspace improve over \textsc{Loud}\xspace, that is, the Anomaly ranker used as a standalone component?
\end{enumerate}

\medskip 

\textit{RQ1} studies the ability of \textsc{Prevent}\xspace to predict failures.  We consider \textcolor{black}{both \textsc{Prevent$_A$}\xspace and \textsc{Prevent$_E$}\xspace,} and comparatively evaluate their ability  to predict failures in terms of false alarm rate, prediction earliness, and stability of true predictions.

\textcolor{black}{We compute the false alarm rate in terms of \emph{false-prediction} and \emph{false-location} alarms.  The \emph{false-prediction alarms} are the timestamps that correspond to states wrongly identified as anomalous during normal execution. 
The \emph{false-location alarms} are the timestamps that correspond to states that are identified as anomalous, but with failures wrongly located in non-anomalous nodes, during failing execution. 
The lower the false alarm rate of either types is, the higher the reliability of the prediction is.}

\textcolor{black}{We compute prediction earliness as the number of timestamps between the first true prediction and the observed system failure, that is, the time interval for activating a healing action before a system failure.
We compute the stability as the true positive rate after the first prediction, that is, the ratio between predictions after the first true prediction and timestamps before the observed system failure. 
Intuitively, the stability indicates the continuity in correctly reporting the failing component in the presence of error states.}

\textit{RQ2} investigates the advantages and limitations of training without failing executions. 
\textsc{Prevent}\xspace trains models with data from normal (non-failing) executions only, while \emph{supervised} techniques rely on training with both normal and failing executions.
The non-necessity of training with failing execution extends the applicability of \textsc{Prevent}\xspace 
with respect to
\emph{supervised} techniques. 
\textsc{Prevent}\xspace discriminates failure-prone conditions as executions that significantly differ from observations at training time, while \emph{supervised} techniques can identify only failures of types considered during training. 
As a result, \textsc{Prevent}\xspace is failure-type agnostic, that is, it can predict failures of any types, 
while \emph{supervised} techniques foster effectiveness by limiting the focus on specific types of failures.
We answer \textit{RQ2} by comparing both \textsc{Prevent$_A$}\xspace and \textsc{Prevent$_E$}\xspace
with \textsc{PreMiSe}\xspace, a state-of-the-art supervised approach that we studied in the last years and successfully applied in industrial settings~\cite{Mariani:PreMiSE:JSS:2020}, in terms of true positive, true negative, and false alarm rates. 

\textcolor{black}{A comparison with state-of-the-art approaches requires a conceivable effort to set up the experimental context and execute the experiments.  
By comparing \textsc{Prevent}\xspace with \textsc{PreMiSe}\xspace we offer a fair comparative evaluation while keeping the costs within an acceptable threshold.
The original \textsc{PreMiSe}\xspace paper~\cite{Mariani:PreMiSE:JSS:2020} reports a comparative evaluation with both \emph{Operation Analytics - Predictive Insights (OA-PI)\footnote{\href{https://www.ibm.com/support/knowledgecenter/en/SSJQQ3_1.3.3/com.ibm.scapi.doc/kc_welcome-scapi.html}{IBM. Operation Analytics - Predictive Insights} Last access: July 2019}}, 
a widely adopted industrial anomaly-based tool, and G-BDA~\cite{Sauvanaud:FaultLocalizationClearwater:ISSRE:2016}, a state-of-the-art signature-based approach, both outperformed by \textsc{PreMiSe}\xspace. Thus, the comparison with \textsc{PreMiSe}\xspace offers a good data spectrum, within acceptable set-up costs because 
of our full access to \textsc{PreMiSe}\xspace. 
}

\textit{RQ3} evaluates the contribution of the synergetic combination of the State classifier and Anomaly ranker, to skim the false alarms that derive from ranking anomalous KPIs in either non-anomalous states or non-anomalous components.
We evaluate the contribution of combining State classifier and Anomaly ranker by comparing both \textsc{Prevent$_A$}\xspace and \textsc{Prevent$_E$}\xspace with \textsc{Loud}\xspace, the Anomaly ranker component standalone that locates faults failures for any state, by referring to the improved approach that we integrated in the \textsc{Prevent}\xspace Anomaly ranker.

\subsection{Experimental Settings}

\begin{figure}[t!]
\begin{center}
\includegraphics[width=75mm]{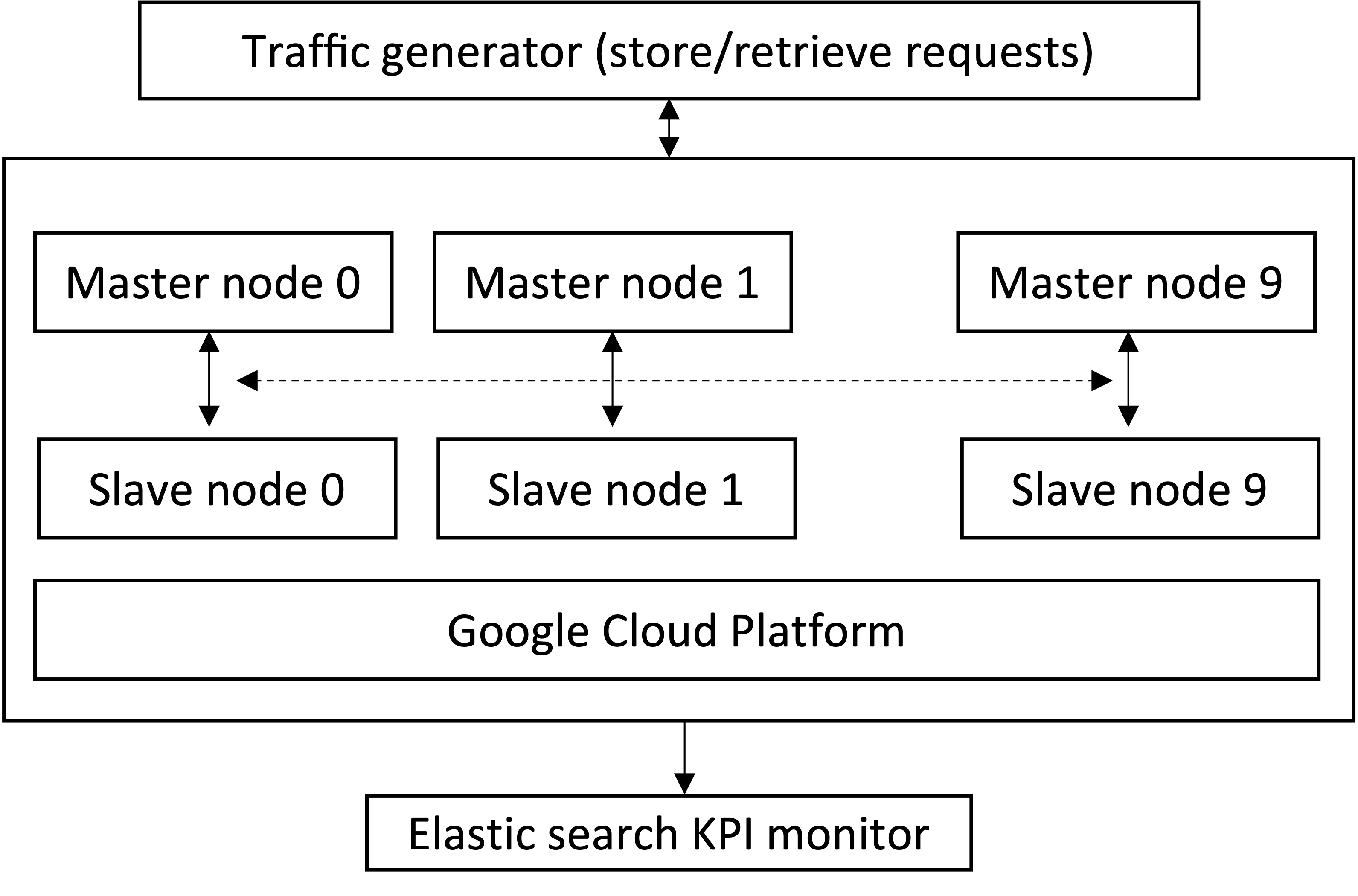}
\caption{Structure of the Redis Cluster 
}
\label{fig:redis}
\end{center}
\end{figure}

We report the results of experimenting with \textsc{Prevent}\xspace on Redis Cluster, a popular open-source enterprise application that provides in-memory data storage.\footnote{\url{https://redis.io}}
A Redis Cluster deploys the  in-memory data storage services on a cluster of computational nodes, and balances the request workload across the nodes to improve throughput and response time.
Figure~\ref{fig:redis} illustrates the structure of the Redis Cluster that we deployed for our experiments: twenty computational nodes running on separate virtual machines, combined pairwise as ten master and ten slave nodes.  We integrated the Redis Cluster with a monitoring facility built on top of Elasticsearch~\cite{elasticsearch:elastic.co}, to collect the 85 KPIs indicated in Table~\ref{fig:kpis}
for each of the twenty nodes of the cluster, on a per-minute basis, resulting in 1,700 KPIs collected per minute.\footnote{The monitoring infrastructure is commonly part of the deployment of distributed applications like Redis, and executing \textsc{Prevent}\xspace at runtime has negligible overhead (less than 1 second) every minute.}

\begin{table*}
\caption{KPIs collected at each node of the Redis cluster}
\label{fig:kpis}
\begin{footnotesize}
\noindent
\begin{tabular}{p{.048\textwidth}p{.03\textwidth} p{.85\textwidth} }
KPI type & \# of KPI & KPI names \\  \hline \hline

Core   & 9 & Id (CPU Core Identifier), idle.pct (Percentage of Idle time), iowait.pct (Percentage of CPU time spent in wait), irq.pct (Percentage of CPU time spent in handling hardware interrupts), nice.pct (Percentage of CPU time spent in low-priority processes), softirq.pct (Percentage of CPU time spent in handling software interrupts), steal.pct (Percentage of CPU time spent in involuntary wait by the virtual CPU while the hypervisor was servicing another processor), system.pct (Percentage of CPU time spent in kernel space), user.pct (Percentage of CPU time spent in user space) \\ \hline

CPU    & 19 & cores (Number of CPU cores on the host), idle.pct, iowait.pct,  irq.pct, nice.pct,  softirq.pct,  steal.pct,  system.pct, total.pct (Percentage of CPU time spent in states other than Idle and IOWait), user.pct
norm.idle.pct, norm.iowait.pct,  norm.irq.pct, norm.nice.pct,  norm.softirq.pct,  norm.steal.pct, norm.system.pct, norm.total.pct , norm.user.pct (norm prefix means normalized value of the corresponding CPU metric without the norm prefix. They are normalized by considering number of cores)
 \\ \hline

File    & 5 & count (Number of file systems), total files (Total number of files), total size.free (Total free space on disk), total size.total (Total disk space either used or free), total size.used (Total used disk space) \\ \hline

Load    & 7 & load.1 (Load average for the last minute), load.5 (Load average for the last 5 minutes), load.15 (Load average for the last 15 minutes), cores (Number of CPU cores on the host), norm.1 (Load in the last minute divided by  number of cores), norm.15, norm.5 \\ \hline

Memory   & 18 &  actual.free, actual.used.bytes, actual.used.pct, free, hugepages.default size, hugepages.free, hugepages.reserved, hugepages.total, hugepages.used.pct, swap.used.bytes, swap.used.pct, total, used.bytes, used.pct \\ \hline

Process   & 8 & dead (Number of dead processes), idle (Number of idle processes), running (Number of running processes), sleeping (Number of sleeping processes), stopped (Number of stopped processes), total (Total number of processes), un-known (Number of processes for which the state could not be retrieved or is unknown), zombie (Number of zombie processes) \\ \hline

Socket   &  11 & all.count (All open connections), all.listening (All listening ports), tcp.all.close wait (Number of TCP connections in close\_wait state), tcp.all.count (All open TCP connections), tcp.all.established (Number of established TCP connections), tcp.all.listening (All TCP listening ports), tcp.all.orphan (Number of all orphaned tcp sockets), tcp.all.time wait (Number of TCP connections in time\_wait state), tcp.memory (Memory used by TCP sockets), udp.all.count (All open UDP connections), udp.memory (Memory used by UDP sockets) \\ \hline

Network & 8 & in.bytes (Number of bytes received), in.dropped (Number of incoming packets that were dropped), in.errors (Number of errors while receiving), in.packets (Number or packets received), out.bytes (Number of bytes sent), out.dropped (Number of outgoing packets that were dropped), out.errors (Number of errors while sending), out.packets (Number or packets received)\\ \hline
\hline
Total & \multicolumn{2}{l}{85 KPI/node (1,700 KPI in the Redis Cluster with 20 nodes)} \\\hline\hline 
\end{tabular}
\end{footnotesize}
\end{table*}

We executed Redis Cluster \textcolor{black}{on Google cloud} with a workload consisting of calls to operations for both storing and retrieving data into and from the Cluster.   
The synthetic workload implements a typical workload of applications in a working environment, with a high amount of calls at daytime and a decreased workload at nighttime, peaks at 7 PM and 9 AM, low traffic in weekends and high traffic in workdays: between 0 and 26 requests per second in the weekends, between 0 and 40  in workdays~\cite{Mariani:LOUD:ICST:2018}.

\subsection{Injecting failures}
We executed the Redis cluster with either no or injected failures, to collect data during normal executions (\emph{normal execution data}) and in the presence of failures (\emph{failing execution data}), respectively.  
\textcolor{black}{We reinitialized Redis and restarted the virtual machines before each execution, and dropped the first 15 timestamps (15 minutes) of execution under normal conditions, to collect data on completely separated runs and experiment in a stable and unbiased context, respectively.  
We injected failures after 15 timestamps of execution under normal conditions. In this way, we reproduce the occurrence of failures in production at any time in normal execution conditions.  
We injected a failure type at a time, to reproduce failures that appear rarely in production.}

\textcolor{black}{We injected failures with \textbf{Chaos Mesh}\footnote{\href{https://chaos-mesh.org/}{https://chaos-mesh.org/}}, a popular tool for injecting failures in the cloud. We experimented with five\xspace of the most common types of failures as defined in  \textbf{Chaos Mesh}: \emph{CPU stress}, \emph{memory stress}, \emph{network packet loss}, \emph{network packet delay}, and \emph{network packet corruption}.}

We executed the Redis cluster with no injected failures for a total of three weeks. We used two weeks of continuous execution \textcolor{black}{ for training (80\%) and validation (20\%). We used the third week of execution, disjoint from the previous two weeks, to collect independent data to check for false positives.}
\textcolor{black}{We injected each failure at a master-slave node pair of the Redis Cluster, and repeated each experiment three times, with failures injected in three different master-slave pairs to reduce statistical biases. }

\bigskip
\subsection*{Experimenting with \textsc{Prevent}\xspace}

\textcolor{black}{We trained the \emph{Deep autoencoder} and the \emph{RBM neural network}, and we built the granger-causality graph for the \emph{granger-causality analyzer} with the unlabeled data collected in two weeks of normal execution. }
\textcolor{black}{We preprocessed the collected KPIs, and discard the ones that kept a stable value all the time, which we identified as the ones for which the variance is below $10^{-5}$ for the entire time series, since these KPIs are not representative of the how the application reacts to input stimuli. We trained our models by considering 719 KPIs out of the 1,700 KPIs initially collected.}

We used the \textcolor{black}{data collected during a third week of normal execution} to
evaluate the impact of \emph{false-prediction alarms}, that is, timestamps that the prediction approaches might erroneously indicate as anomalous states during normal executions.

We used the failing execution data to evaluate 
\begin{inparaenum}[(i)]
\item the ability of revealing failures,
\item the ability to correctly locate the corresponding failing nodes, 
\textcolor{black}{\item the impact of both \emph{false prediction} and \emph{false location alarms},
that is, the timestamps that correspond to states wrongly identified as anomalous during normal executions, and the timestamps that correspond to wrongly identified localizations during failing executions, respectively, as defined in Section~\ref{sec:approach}.}
\end{inparaenum}

\subsection*{Experimenting with \textsc{PreMiSe}\xspace and \textsc{Loud}\xspace}

The supervised \textsc{PreMiSe}\xspace approach requires training with labeled data.  
To train \textsc{PreMiSe}\xspace, we labeled normal execution data as \emph{no-failure}, and we labelled the failing execution data as $\langle \emph{failure type}, \emph{master-slave node pair}\rangle$ to indicate the type of the injected failure and the pair of nodes at which the failures were injected, respectively. 
\textcolor{black}{We augmented the training data set to include failing execution data
for all nodes, synthesized by replicating failures on symmetric nodes. 
In this way, we obtained ten failing data sets for each original failing data set, one per node pair. 
For each failure type $FT$, we trained \textsc{PreMiSe}\xspace with the labeled data from both normal and failing executions, without the data from the failing executions corresponding to $FT$, and we evaluated the ability of \textsc{PreMiSe}\xspace to predict failures when induced by faults of type $FT$.}

We compared \textsc{Prevent}\xspace with \textsc{Loud}\xspace, by considering the same embodiment of \textsc{Loud}\xspace
that we exploit in \textsc{Prevent}\xspace, that is, the standalone Anomaly ranker component trained as described above. 

\textsc{Loud}\xspace (the standalone Anomaly ranker) ranks anomalous KPIs for all execution states, regardless of any prediction, and thus suffers from many false-prediction alarms on normal execution data.  
For a fair comparison, we filtered out false-prediction alarms by signaling failure predictions only when \textcolor{black}{\textsc{Loud}\xspace reports the same}  \textcolor{black}{node at the top of the anomalous node ranking} for $N$ consecutive timestamps. This follows the intuition that anomalies increase in the presence of failures, and thus the failing nodes should likely persist as the top ranked nodes if the failing conditions keep occurring.
We report the false-prediction alarm rate of \textsc{Loud}\xspace for $N=3,4,5,6$, after observing a huge amount of false alarms for  values of $N$ less than 3, and an unacceptable delay in signaling anomalous states for values of $N$ greater than 6.

\subsection{Results}

\begin{figure}[!htb]
\begin{center}
\includegraphics[width=90mm]{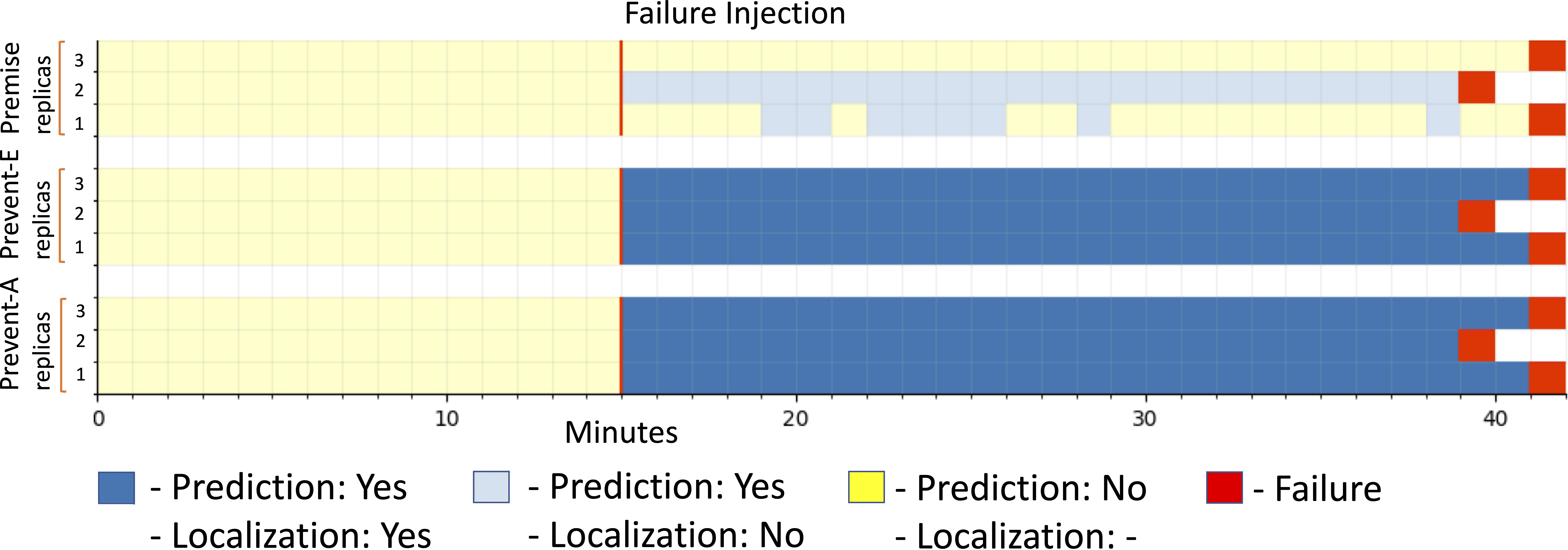} 
\end{center}
\caption{Failure predictions for CPU stress}
\label{fig:cpuhog}
\end{figure}

\begin{figure}[!htb]
\begin{center}
\includegraphics[width=85mm]{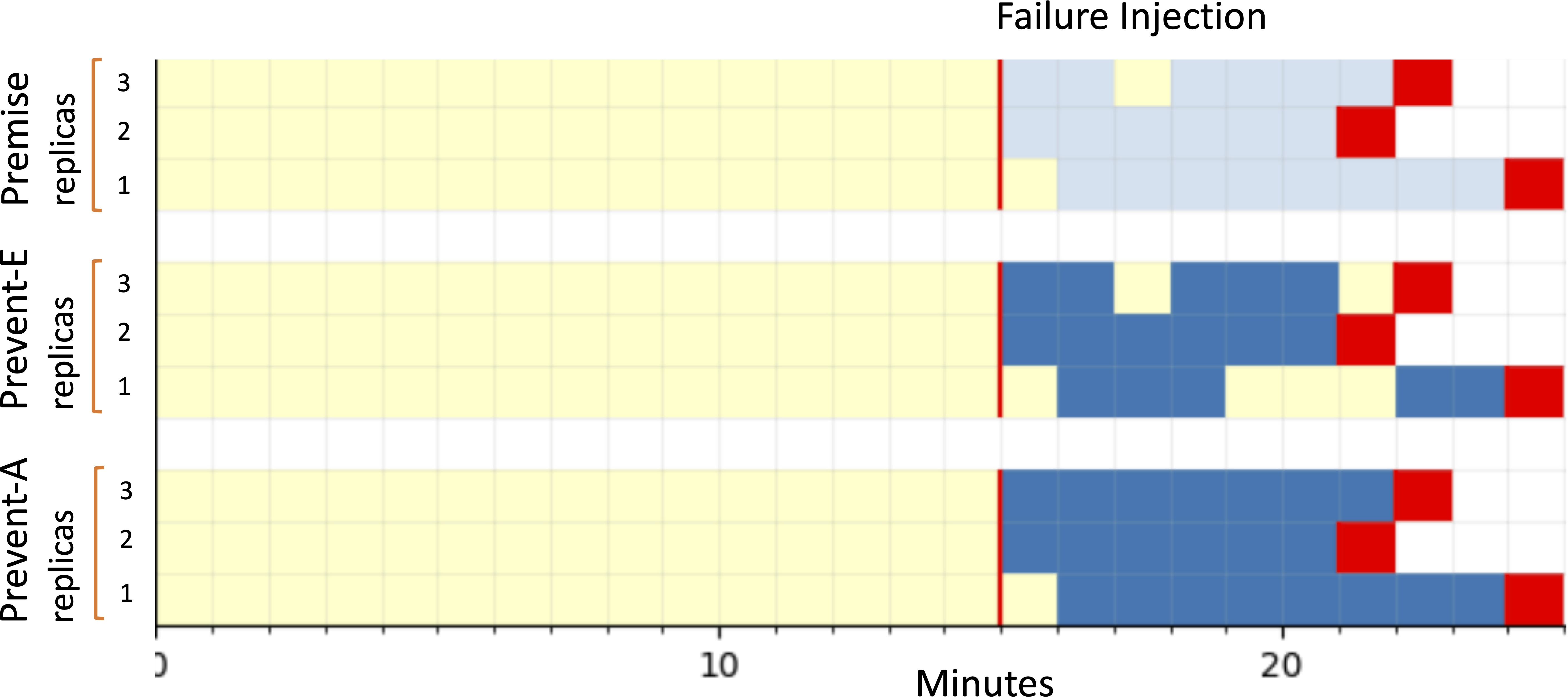} 
\end{center}
\caption{Failure predictions for packet loss}
\label{fig:packetloss}
\end{figure}

\begin{figure}[!htb]
\begin{center}
\includegraphics[width=85mm]{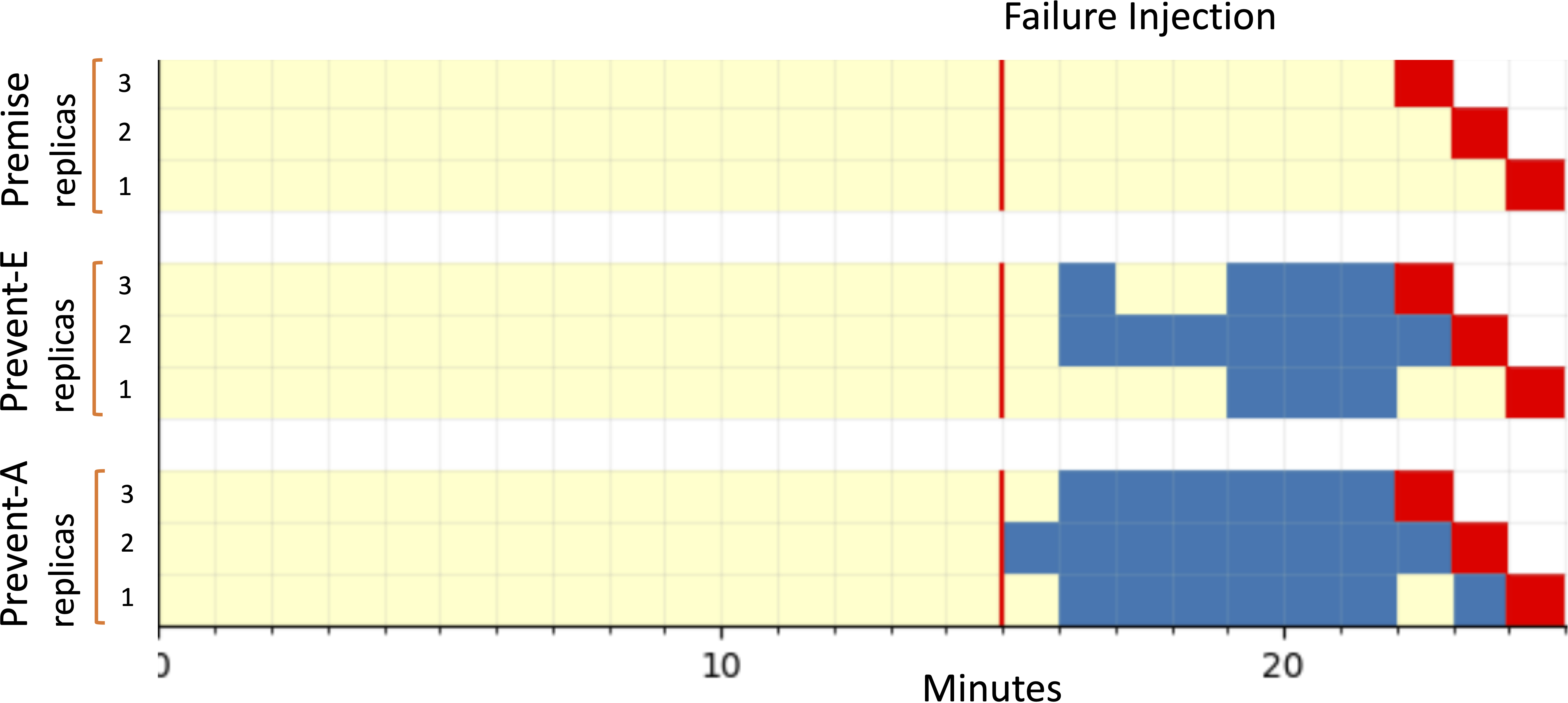} 
\end{center}
\caption{Failure predictions for packet delay}
\label{fig:packetdelay}
\end{figure}

\begin{figure}[!htb]
\begin{center}
\includegraphics[width=85mm]{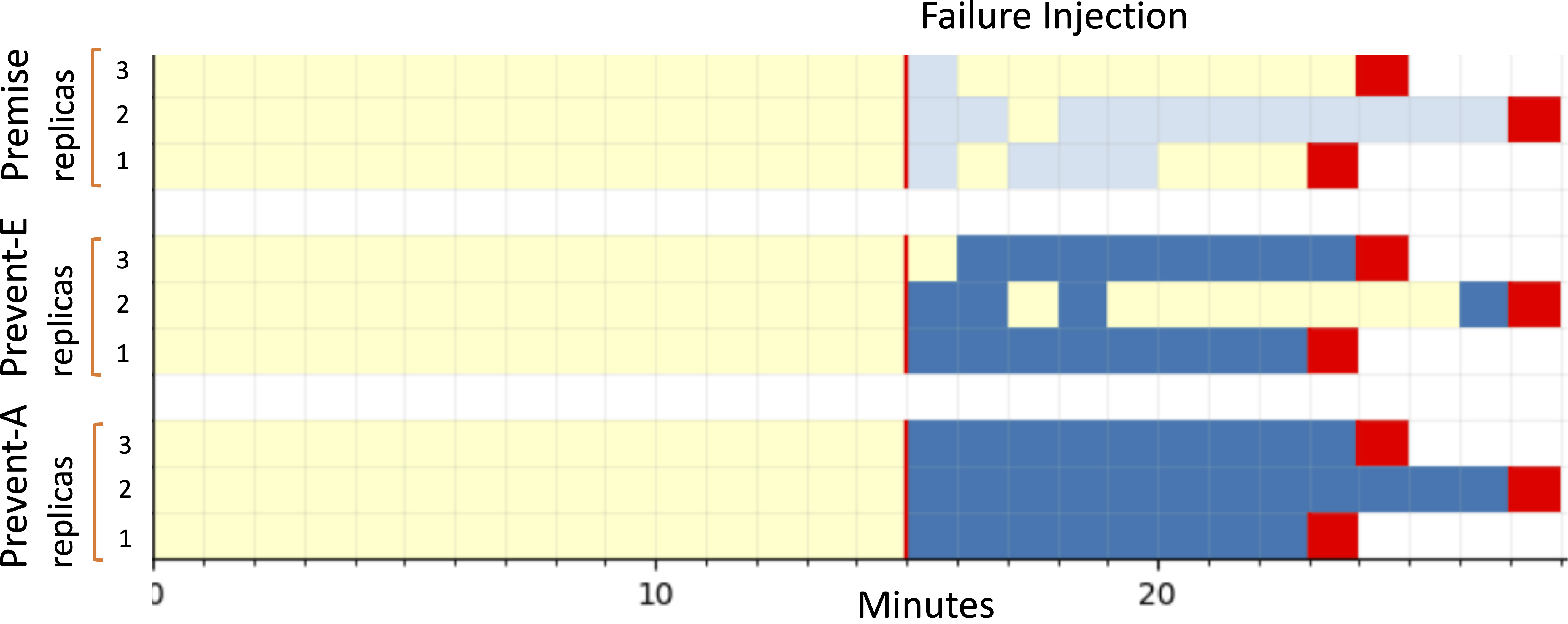} 
\end{center}
\caption{Failure predictions for packet corruption}
\label{fig:packetcorruption}
\end{figure}

Figures~\ref{fig:cpuhog},~\ref{fig:memoryleak},~\ref{fig:packetloss},~\ref{fig:packetdelay},~\ref{fig:packetcorruption}
visualize the results of \textsc{Prevent$_A$}\xspace, \textsc{Prevent$_E$}\xspace, and \textsc{PreMiSe}\xspace in the experiments with the injected failures. 
Each figure shows three plots that report the results of three replicas of the experiments for \textsc{Prevent$_A$}\xspace, \textsc{Prevent$_E$}\xspace, and \textsc{PreMiSe}\xspace, respectively.
\textcolor{black}{The plots of \textsc{PreMiSe}\xspace report the data obtained with}
the \textsc{PreMiSe}\xspace model \textcolor{black}{trained without the failures of the type considered in the plot.}

\begin{figure}[!htb]
\begin{center}
\includegraphics[width=90mm]{imgs/mem-leak} 
\end{center}
\caption{Failure predictions for memory leak}
\label{fig:memoryleak}
\end{figure}

The x-axis indicates the timeline of each experiment \textcolor{black}{ in minutes: 
\textcolor{black}{Each experiment includes a (not showed)} initial 15~minutes lag to stabilize the system, a fault injection \textcolor{black}{after additional 15 minutes of normal execution (the vertical red line in the plots)}, and a failing execution (the area after the red line) up to an observable system failure (a red square at the end of the plots). 
}

\textcolor{black}{We observe a Redis failures when the memory fragmentation ratio of Redis servers exceeds 1.5, following the Redis documentation that requires the memory fragmentation ratio of Redis servers to be lower than 1.5, and the servers to be restarted otherwise.\footnote{\url{https://redis.io/docs}}}

\textcolor{black}{Each plot
shows three rows of colored squares, one row for each
experiment replica.}
\textcolor{black}{The colors}  
visualize the strength of the predictions of the different approaches at each timestamp:
\begin{itemize}
\item \textcolor{black}{Blue squares indicate successful failure predictions, that is, failure predictions reported along with correct localization results. 
In the case of \textsc{Prevent}\xspace this means that the node reported at the top of the localization ranking is a node in which we injected the failure.}

\item \textcolor{black}{Grey squares indicate either false prediction alarms before the injection or false location alarms after the injection, that is, states wrongly identified as anomalous during normal executions and failures wrongly located in non-anomalous nodes, respectively.
}

\item \textcolor{black}{Yellow squares indicate the absence of prediction, which correspond to either true negatives before the start of the failure injection or false negatives after the injection.}

\item \textcolor{black}{The red squares indicate the timestamps at which the injected failures manifest as observable failures of the cluster, and the experiment terminates.}
\end{itemize}

\textcolor{black}{We add the legend only to Figure~\ref{fig:cpuhog} to reduce redundancy. 
In a nutshell, the more the yellow squares occur before the injection and the blue squares occur after the injection, the better the approach is.
As we discuss below, the three experiment replicas yields consistent results in all cases, thus suggesting that three replicas are sufficient indeed.}

\medskip

\medskip

\subsection*{RQ1: Effectiveness}

RQ1 focuses on \textsc{Prevent$_A$}\xspace and \textsc{Prevent$_E$}\xspace  \textcolor{black}{(the bottom  and mid plots} in Figures~\ref{fig:cpuhog},~\ref{fig:memoryleak},~\ref{fig:packetloss},~\ref{fig:packetdelay},~\ref{fig:packetcorruption}).
We comparatively evaluate the effectiveness of \textsc{Prevent$_A$}\xspace and \textsc{Prevent$_E$}\xspace to predict failures in terms of false prediction alarm rate, 
reaction time, prediction earliness and true positive rate that we define as follows:
\begin{itemize}
\item \textcolor{black}{\emph{False prediction alarm rate:}  The portion of false predictions during normal execution, before the failure injection, that is, the portion of grey squares before the vertical red line;}
\item \emph{Reaction time:} The time interval between the injection and the first true prediction, that is, the number of squares between the failure injection and the first blue square; 
\item \emph{Prediction earliness:} \textcolor{black}{The time interval between the first true positive and the observable system failures, that is, the number of squares between the first blue square and the red square;}
\item \emph{True positive rate:} The portion of true positives between the first true positive and the  observable system failures, that is, the portion of \textcolor{black}{blue squares}  between the first \textcolor{black}{blue square} and the \textcolor{black}{red square}. 
\end{itemize}

The \textcolor{black}{false prediction} alarm rate quantifies the annoyance of operators receiving useless alerts.  
The reaction time quantifies the delay in alerting the operators.
The prediction earliness quantifies the time offered to the operators to adopt countermeasures before the observable system failure. 
The true positive rate indicates the stability of the \textcolor{black}{failure location alarms, once they are first raised.}
Predictions with high stability \textcolor{black}{are most desirable since they indicate good sensitivity of the models after they start sensing the failure symptoms.} 

The plots in the figures show that both \textsc{Prevent$_A$}\xspace (the bottom plots) and \textsc{Prevent$_E$}\xspace (the mid plots) 
successfully predict all five types of failures, consistently across the experiment replicas. 
\textsc{Prevent$_A$}\xspace predicts both slightly earlier and with slightly better stability than \textsc{Prevent$_E$}\xspace  both packet delay and packet corruption failures.
\textcolor{black}{The yellow squares before the injection of the failure indicate that both  \textsc{Prevent$_A$}\xspace and \textsc{Prevent$_E$}\xspace
 do not suffer from false-prediction alarms during normal execution.}

Tables~\ref{tab:earliness:anomaly} and~\ref{tab:earliness:energy} report in detail the reaction time interval, the prediction earliness interval, and the true positive rate of predictions (TPR) per failure type \textcolor{black}{and experiment replica,} for \textsc{Prevent$_A$}\xspace and \textsc{Prevent$_E$}\xspace, respectively.
The data indicate that both \textsc{Prevent$_A$}\xspace and \textsc{Prevent$_E$}\xspace predict failures
\textcolor{black}{within the first minute after the failure injection in most cases (reaction interval $\leq 1$), and in four minutes in the worst case. 
In details,  \textsc{Prevent$_E$}\xspace predicts the failure in less than a minute in 13 out of 15 cases, and  \textsc{Prevent$_A$}\xspace in 12 out of 15 cases.
 The predictions of \textsc{Prevent$_A$}\xspace are more stable than the predictions of \textsc{Prevent$_E$}\xspace for packet loss, packet delay and packet corruption. 
The TPR of \textsc{Prevent$_A$}\xspace is 100\% in 14 out of 15 cases, and 88\% in a single replica of the packet delay experiment, while the TPR of \textsc{Prevent$_E$}\xspace is 100\% in 10 cases; it is between 60\% and 71\% in 4 cases; It is 33\% in a replica of the packet corruption experiment.}

Overall  both \textsc{Prevent$_A$}\xspace and \textsc{Prevent$_E$}\xspace predict failures early and with good stability in most cases. 
\textcolor{black}{The two approaches have comparable performance, although \textsc{Prevent$_A$}\xspace provides slightly more stable predictions than \textsc{Prevent$_E$}\xspace.
These results confirm the ability of the general \textsc{Prevent}\xspace approach to predict failures in unsupervised fashion, and the relevance of the original  \textsc{Prevent$_A$}\xspace instantiation that we propose in this paper.}

\begin{table}[!htb]
\caption{Prediction earliness for \textsc{Prevent$_A$}\xspace}
\label{tab:earliness:anomaly}
\begin{footnotesize}

\textcolor{black}{
\begin{tabular}{p{0.07\textwidth}|p{0.07\textwidth}|p{0.05\textwidth}|p{0.05\textwidth}|p{0.05\textwidth}}
\hline
Failure Type & Experiment replica & reaction interval & earliness interval & TPR \\
&  \multicolumn{3}{c}{(in minutes)}  & \\
\hline \hline
\multirow{3}{*}{CPU stress} 
	& r1 & 0 & 26 &  100\%\\
 	& r2 & 0 &  24 & 100\% \\
 	& r3 & 0 &  26 & 100\% \\ \hline
\multirow{3}{*}{Mem leak}	
	& r1  & 2 & 33 &  100\%\\
	& r2 & 2 & 25 & 100\% \\
	& r3 & 3 & 30 & 100\% \\\hline
\multirow{3}{*}{Pckt loss}
	& r1 & 1 & 8 & 100\%\\
	& r2 & 0 &  6 &  100\% \\
	& r3 & 0 &  7 &  100\% \\ \hline
\multirow{3}{*}{Pckt delay}
	& r1 & 1 & 8 & 88\%\\
	& r2 & 0 & 8 & 100\%\\
	& r3 & 1 & 6 & 100\%\\\hline
\multirow{3}{*}{Pckt corr}
	& r1 & 0 & 8 & 100\%\\
	& r2 & 0 & 12 & 100\%\\
	& r3 & 0 & 9 & 100\%\\
\hline \hline
\end{tabular} 
}

\end{footnotesize}
\end{table}

\begin{table}[!htb]
\caption{Prediction earliness for \textsc{Prevent$_E$}\xspace}
\label{tab:earliness:energy}
\begin{footnotesize}

\textcolor{black}{
\begin{tabular}{p{0.07\textwidth}|p{0.07\textwidth}|p{0.05\textwidth}|p{0.05\textwidth}|p{0.05\textwidth}}
\hline
Failure Type& Experiment replica & reaction interval & earliness interval & TPR \\
&  \multicolumn{3}{c}{(in minutes)} &\\
\hline \hline
\multirow{3}{*}{CPU stress} 
	& r1  & 0 & 26 &  100\%\\
 	& r2  & 0 &  24 & 100\% \\
 	& r3  & 0 &  26 & 100\% \\ \hline
\multirow{3}{*}{Mem leak}	
	& r1 & 1 & 34 &  100\%\\
	& r2 & 2 & 25 & 100\% \\
	& r3 & 1 & 32 & 100\% \\\hline
\multirow{3}{*}{Pckt loss}
	& r1 & 1 & 8 & 63\%\\
	& r2 & 0 &  6 &  100\% \\
	& r3 & 0 &  7 &  71\% \\ \hline
\multirow{3}{*}{Pckt delay}
	& r1 & 4 & 5 & 60\%\\
	& r2 & 1 & 7 & 100\%\\
	& r3 & 1 & 6 & 67\%\\\hline
\multirow{3}{*}{Pckt corr}
	& r1 & 0 & 8 & 100\%\\
	& r2 & 0 & 12 & 33\%\\
	& r3 & 1 & 8 & 100\%\\
\hline \hline
\end{tabular} 
}

\end{footnotesize}
\end{table}

\subsection*{RQ2: Comparative Evaluation}

RQ2 focuses on the advantages and limitations of the unsupervised \textsc{Prevent}\xspace approach with respect to state-of-the-art (supervised) approaches.   Unsupervised approaches do not require training with data collected during failing executions, thus they can be used in the many industrially relevant cases where it is not possible to seed failures during operations, as required for supervised approaches.

\textcolor{black}{We experimentally compare \textsc{Prevent}\xspace with \textsc{PreMiSe}\xspace, a representative supervised state-of-the-art approach as we discuss in Section~\ref{sec:approach} at page~\pageref{line:premise}}, to understand the impact of the lack of training with data from failing execution on earliness and true positive rate.   
The top plots in Figures~\ref{fig:cpuhog},~\ref{fig:memoryleak},~\ref{fig:packetloss},~\ref{fig:packetdelay},~\ref{fig:packetcorruption} visualize the results  of executing \textsc{PreMiSe}\xspace on the same data we use in the experiments with \textsc{Prevent$_A$}\xspace and \textsc{Prevent$_E$}\xspace.

 \textcolor{black}{In the experiments \textsc{PreMiSe}\xspace rarely predicts failures and never localizes the faulty component (grey squares after failure injection, that is, false location alarms).}
 The experiments confirm that the original combination of a state classifier with an anomaly ranker of both \textsc{Prevent$_A$}\xspace and \textsc{Prevent$_E$}\xspace is more effective than the \textsc{PreMiSe}\xspace supervised learning approach, in terms of both true and false positive rate.

\begin{footnotesize}

\end{footnotesize}

\subsection*{RQ3: state classifier and anomaly ranker interplay}

RQ3 focuses on the effectiveness of the original combination of an anomaly ranker with a state classifier in both \textsc{Prevent$_A$}\xspace and \textsc{Prevent$_E$}\xspace.  We compare \textsc{Prevent$_A$}\xspace and \textsc{Prevent$_E$}\xspace with \textsc{Loud}\xspace, the state-of-the-art anomaly ranker that inspired our anomaly ranker.  
We estimated the false-prediction alarm rate of \textsc{Prevent$_A$}\xspace, \textsc{Prevent$_E$}\xspace,
and \textsc{Loud}\xspace (with $N=3,4,5,6$) by executing each approach with the data that we collected during a week of normal execution of the Redis Cluster \textcolor{black}{(the third week of our collected data, disjoint from the two weeks of data that we used for training)}.

Table~\ref{tab:false-alarms-normal} reports the \textcolor{black}{false-prediction alarm rate, that is, the portion of false predictions during the week of normal data.}
The false-prediction alarm rate quantifies the annoyance of operators receiving useless alerts.
\textsc{Prevent$_A$}\xspace and \textsc{Prevent$_E$}\xspace dramatically reduced the false-prediction alarm rate from unacceptable values between 22\% and 57\%, when using \textsc{Loud}\xspace stand-alone, to 2\% and 5\%, respectively, by combining \textsc{Loud}\xspace with failure predictors, as in \textsc{Prevent}\xspace. \textcolor{black}{\textsc{Prevent$_A$}\xspace outperforms all approaches.}

\begin{table}
\centering
\caption{False alarm rate 
on the
normal execution
data}
\label{tab:false-alarms-normal}
\begin{footnotesize}
\begin{tabular}{p{0.07\textwidth}p{0.18\textwidth}}
\hline
Approach & False-prediction alarm rate \\
\hline \hline
\textsc{Prevent$_A$}\xspace & 2\%\\
\textsc{Prevent$_E$}\xspace & 5\%   \\
\textsc{Loud}\xspace{$N3$} & 57\% \\
\textsc{Loud}\xspace{$N4$} & 43\%\\
\textsc{Loud}\xspace{$N5$} & 32\%\\
\textsc{Loud}\xspace{$N6$} & 22\%\\
\hline \hline
\end{tabular} 
\end{footnotesize}
\end{table}

\medskip

\subsection*{Threats to validity}

We evaluated \textsc{Prevent}\xspace by executing a prototype implementation on a large dataset that we collected on a complex cluster. We carefully implemented the approach and carefully designed our experiments to avoid biases.  We understand that the limited experimental framework may threaten the validity of the results, and we would like to conclude this section by discussing how we mitigated the main threats to the validity of the experiments that we identified in our work. 

\bigskip
\noindent
\emph{Threats to the external validity}\\

Possible threats to the external validity of the experiments derive from the  experimental evidence being available for a single system and 
 \textcolor{black}{five failure types}, which in turn depend on the effort required to set up a proper experimental environment and collect data for validating the approach.  The main threats derive from the experimental setting, and the set of failures.     

\bigskip
\noindent
\emph{Experimental setting}\\
We experimented with a dataset collected from a Redis cluster that we implemented on the Google Cloud Platform, controlling the stability of the cluster and guaranteeing the same running conditions for all experiments, and the results may not generalise to other systems.
We mitigated the threat that derive from experimenting with a single system by collecting data from a standard installation of a popular distributed application that is widely used in commercial environments, and by collecting a large set of data from several weeks of experiments.  
The data we used in the experiments are available in a replication package\footnote{The replication package at~\href{https://star.inf.usi.ch/\#/software-data/14}{https://star.inf.usi.ch/\#/software-data/14}\@\xspace} for replicating the results and for comparative studies with new systems.  

\bigskip
\noindent
\emph{Set of failures}\\
We experimented with \textcolor{black}{five} failure types, and the results may not generalize to other failure types. 
We limited the experiments to \textcolor{black}{five} failure types due to the large effort required to properly install and tune failure injectors and collect valid data sets for each failure.  We mitigated this threat to the validity of our results by choosing types of failures which are both common and very relevant in complex cloud systems, \textcolor{black}{and by injecting the failures with \textbf{Chaos Mesh}\footnote{\href{https://chaos-mesh.org/}{https://chaos-mesh.org/}}, a failures injector commonly used to study failure in cloud systems.}

\bigskip
\noindent
\emph{Threats to the internal validity}\\

Possible threats to the internal validity of the experiments derive from the prototype implementation of the approaches and the traffic profile used in the experiments.

\bigskip
\noindent
\emph{Prototype implementation}\\
We carried on the experiments on an in-house prototype implementation of \textsc{Prevent}\xspace, \textsc{PreMiSe}\xspace and \textsc{Loud}\xspace. We mitigated the threats to the validity of the experiments that may derive from a faulty implementation by carefully designing and testing our implementation, by 
\begin{inparaenum}[(i)]
\item comparing the data obtained with our \textsc{PreMiSe}\xspace and \textsc{Loud}\xspace implementations with the data available in the literature, 
\item repeating the experiments three times, and 
\item making the prototype implementation available in a replication package, to allow for replicating the results.
\end{inparaenum}

\bigskip
\noindent
\emph{Traffic profile}\\
We mitigated the threats to the validity of the results that may derive from the traffic profile that we use to collect the experimental data, by experimenting with a seasonal traffic profile excerpted from industrial collaborations, and that we compared with analogous traffic profiles publicly available in the literature.  We carried on the experiments by collecting data over weeks of continuous executions with profiles that correspond to traffic commonly observed in commercial settings.

\section{Related Work} \label{sec:related}

\textcolor{black}{In this section we overview the applications of machine learning to software engineering, and discuss in detail the approaches for predicting software failures at runtime.}

\subsection{\textcolor{black}{Machine Learning for Software Engineering}}

\textcolor{black}{Many studies address software engineering problems with supervised, unsupervised, weakly-supervised and semi-supervised machine learning strategies.}

\textcolor{black}{Supervised learning approaches require training samples annotated (labelled) with the prediction outcomes.
The supervised training phase tunes the parameters of the model by minimizing the errors with respect to the outcomes encoded with the labels. There are many supervised learning approaches for predicting failures~\cite{ozcelik:seer:tse:2016,Malik:AutomaticDetection:ICSE:2013,Sauvanaud:FaultLocalizationClearwater:ISSRE:2016,nistor:suncat:issta:2014,Mariani:PreMiSE:JSS:2020}.}

\textcolor{black}{Unsupervised learning approaches work with unlabelled datasets, and either do not require training at all or tune models that represent the characteristics of the samples as a whole, as in our \textsc{Prevent}\xspace approach. 
In production, unsupervised learning approaches either discriminate clusters of mutually similar samples~\cite{Nam:UnlabeledDefectPrediction:IEEE:2015} or identify anomalies as samples that notably differ from the samples used in the training phase~\cite{monni:RBM:ICST:2019,Fernandes:anomalydetection:JNCA:2016,Bontemps:anomalydetection:FDSE:2016,Du:deeplog:CCS:2017}, as in \textsc{Prevent}\xspace.}
\textcolor{black}{Unsupervised approaches do not require the expensive effort of annotating large training samples, while 
supervised approaches precisely predict the outcomes observed in the training set.} 

\textcolor{black}{Semi-supervised approaches are an interesting tradeoff between costs and precision, and have been recently applied to solve several software engineering problems, including predicting software fault-proneness~\cite{tu:frugal:ase:2021,zhang:lowrank:iet:2021,lu:defect:ase:2012}.}

\textcolor{black}{Weak-supervised approaches aim to achieve the benefits of supervised learning, while reducing the costs of labelling~\cite{Zhang:WeakSupervision:CoRR2022,Bach:Learning:GenerativeModelsWithoutLabeledData:ICML:2017,Liang:LearningFromMeasurements:ICML:2009,Ratner:LargeTrainingSetsQuickly:ANIPS:2016,Pan:TransferLearningSurvey:IEEE:2010,Stewart:LabelFreeSupervision:AAAI:2017}. 
The popular \emph{active learning} approaches require to label a relatively small set of training samples that the active learning algorithms select as the samples that contribute most to improve the quality of the predictions~\cite{Druck:ActiveLearning:EMNLP:2009}.}  

\textcolor{black}{\emph{Semi-supervised learning} approaches are weak-supervised approaches that rely on samples that the approaches  automatically label from the characteristics of a small set of data that come with known labels~\cite{Chapelle:SemiSupervisedLearning:MIT:2006,Berthelot:MixmatchSemiSupervisedLearning:ANIPS:2019,Laine:TemporalEnsemblingSemiSupervisedLearning:arXiv:2016,Mann:SemiSupervisedLearningWeaklyLabeled:MLR:2010}.
Bodo et al. propose a semi-supervised learning approach to analyze the performance indicators related to the software process, and show that the semi-supervised algorithm outperforms supervised learning approaches~\cite{bodo:process:serp:2015}. 
There are many semi-supervised learning approaches for software analytics, and in particular for predicting fault-prone software modules~\cite{tu:frugal:ase:2021,zhang:lowrank:iet:2021,lu:defect:ase:2012}. }

\textcolor{black}{Classic approaches  identify fault-prone modules with supervised models based on metrics that reflect the complexity of the code. These approaches require training datasets labeled either as defective or non-defective according to historical data, with a high cost for collecting the training samples~\cite{denaro:fault-proneness:seke:2002}. 
Zhang et al. alleviate the labelling cost by exploiting the similarity among the code-level metrics of software modules. They propose a semi-supervised learning approach that propagate an initially small set of the labelled samples to unlabelled samples, by applying spectral clustering~\cite{zhang:lowrank:iet:2021}.
Tu and Menzies’~\textsc{Frugal} approach extends the semi-supervised approach from fault-proneness predictions to other software analytics, like code warnings and issue close time~\cite{tu:frugal:ase:2021}.}
  
\textcolor{black}{The problem of identifying fault-prone software from both historical data on the defects identified in the software and code metrics that can be measured during software testing and maintenance radically differs from the problem of predicting failures at runtime from metrics that measure the execution of the software and not the code. To the best of our knowledge, there exist no semi-supervised or weak-supervised approaches to predict failures at runtime.}

\subsection{\textcolor{black}{Failure Prediction and Diagnosis}}

Salfner et al.'s survey identifies \emph{online failure prediction} as the first step to proactively manage faults in production, followed by \emph{diagnosis}, \emph{action scheduling} and \emph{execution of corrective actions},   
and defines \emph{failure symptoms} as the out-of-norm behavior of some system parameters before the occurrence of a failure, due to side effects of the faults that are causing the failure~\cite{Salfner:PredictSurv:ACMCompSurv:2010}.
In this section we focus on the failure prediction and diagnosis steps of Salfner~et~al.'s~fault management process, the steps related to \textsc{Prevent}\xspace.  We refer the interested readers to Colman-Meixner~et~al.'s~comprehensive survey  for a discussion of mechanisms for scheduling and executing corrective actions to tolerate or recover from failures~\cite{colman-meixner:survey-resiliency:comst:2016}.

Detecting failure symptoms, often referred to as \emph{anomalies}, is the most common approach to predict failures online.
The problem of detecting anomalies has been studied in many application domains, to reveal intrusions in computer systems, frauds in commercial organizations, abnormal patient conditions in medical systems, damages in industrial equipments, abnormal elements in images and texts, abnormal events in sensor networks, failing conditions in complex software systems~\cite{Chandola:AnomalyDetetionSurvey:CSUR:2009}. 
Approaches to detect anomalies heavily depend on the characteristics of the application domain.
In this section, we discuss approaches to detect anomalies in complex software systems, to predict failures in production.

A distinctive characteristics of approaches for detecting anomalies is the model that the approaches use to interpret the data monitored in production, models that are either manually derived by software analysts in the form of rules that the system shall satisfy~\cite{Chung:bottleneckdetection:IPDPS:2008} or automatically inferred from the monitored data. 
Approaches that automatically infer models from monitored data  work in either supervised or unsupervised fashion, and do or do not require labeled data to synthesize the models, respectively. 

\emph{Rule-based approaches} leverage analysts' knowledge about the symptoms that characterize failures, and rely on rules manually defined for each application and context.
\emph{Supervised approaches} that are also referred to as \emph{signature-based approaches}~\cite{Ibidunmoye:AnomalyDetectionSurvey:2015} build the models by relying on previously observed anomalies, and offer limited support for detecting anomalies that were not previously observed. 
\emph{Unsupervised approaches} derive models without requiring labeled data, thus better balancing accuracy and required information.
Some approaches focus on failure predictions only, yet others approaches support both prediction and diagnosis, thus addressing the first two steps of  Salfner et al.'s proactive fault management process.

The \textsc{Prevent}\xspace approach that we propose in this paper is an unsupervised approach that both predicts and diagnoses failures, by detecting anomalies in the KPI values monitored on distributed enterprise applications. 
The \textsc{Prevent}\xspace State classifier predicts upcoming failures as it observes system states with significant sets of anomalous KPIs. The \textsc{Prevent}\xspace Anomaly ranker diagnoses the components that correspond to the largest sets of representative anomalies as the likely originators of the failures.  
In the remainder of this section we review the relevant automatic approaches to predict and diagnose faults in complex software systems.

\subsection*{Failure Prediction}

The failure prediction approaches reported in the literature predict failures in either datacenter hosts that serve cloud-based applications and services or distributed applications that span multiple hosts. 

Approaches that predict failures in datacenter hosts monitor metrics about resource consumption at host levels, like CPU, disk, network and memory consumption, to infer whether a host is likely to incur some failure in the near future, and proactively migrate applications that run on likely-failing hosts.
Tehrani and Safi exploit a support-vector-machine model against (discretized) data on CPU usage, network bandwidth and available memory of the hosts, to identify hosts that are about to fail~\cite{tehrani:threshold-sensitive:jmgs:2017}. 
Both Islam and Manivannan and Gao~et~al.~investigate the relationship between resource consumption and task/job failures, to predict failures in cloud applications~\cite{islam:failure-in-cloud:iccc:2017,gao:task-failure:bigdata:2019}.
Both David et al.\  and Sun et al.\ exploit neural networks against data on disk and memory usage, to predict hardware failures, isolate the at-risk hardware, and backup the data~\cite{davis:failuresim:cloud:2017,sun:hardware-failure:dac:2019}.
Approaches that predict failures in distributed applications, like \textsc{Prevent}\xspace,
address the complex challenge of sets of anomalies that span multiple physical and virtual hosts and devices. 

\emph{Signature-based} approaches leverage the information of observed anomalies with supervised learning models, to capture the behavior of the monitored system when affected by specific failures.  
Signature-based approaches are very popular and support a large variety of approaches.
Among the most representative approaches,  both \emph{Seers}~\cite{ozcelik:seer:tse:2016} and  Malik ~et~al.'s~approaches~\cite{Malik:AutomaticDetection:ICSE:2013} label runtime performance data as related to either passing or failing executions, and train classifiers to identify incoming failures by processing runtime data in production,
Sauvanaud et al.\ classify service level agreement violations based on data collected at the architectural tier ~\cite{Sauvanaud:FaultLocalizationClearwater:ISSRE:2016},
The \emph{SunCat} approach models and predicts performance problems in smartphone applications~\cite{nistor:suncat:issta:2014}.

Signature-based approaches suffer from two main limitations: 
 they both require labeled failure data for training, which may not be easy to collect, and 
foster predictions that overfit the failures available in the training set, without well capturing other failures and failure patterns. 

Both \emph{semi-supervised} and \emph{unsupervised} approaches learn models from  data observed during normal (non-failing) executions, and identify anomalies when the data observed in production deviate from the normal behaviors captured in the models, thus discharging the burden of labelling the training data.

Most  approaches that claim a semi-supervised nature combine semi-supervised learning with signature-based models to yield their final prediction verdicts~\cite{Fulp:PredictingFailure:WASL:2008}~\cite{tan:prepare:icdcs:2012}~\cite{Tan:AnomalyPrediction:PODC:2010}~\cite{guan:ensemble:jcom:2012}.
As representative examples, Mariani et al.'s \textsc{PreMiSe}\xspace approach
uses a semi-supervised approach on monitored KPIs (IBM SCAPI~\cite{IBM:SCAPI:Tutorial:2014})
 to identify anomalous KPIs, and builds a signature-based model to identify failure-prone anomalies, by capturing the relation between failures observed in the past and the sets of anomalous KPIs~\cite{Mariani:PreMiSE:JSS:2020}. 
Fulp et al.'s approach builds failure prediction models with a support vector machine (a signature-based approach) based on features distilled in semi-supervised fashion from system log files ~\cite{Fulp:PredictingFailure:WASL:2008}.
 Tan et al.'s \emph{PREPARE} approach predicts anomalies with a Bayesian networks trained in semi-supervised fashion, and feeds the anomalies to a 2-dependent Markov model (a signature based model) to predict failures~\cite{tan:prepare:icdcs:2012}.
Tan et al.'s ALERT approach refers to unsupervised clustering to map the failure data to distinct failure contexts, aiming to  predict  failures by using a distinct decision tree (a signature-based approach) for each failure context~\cite{Tan:AnomalyPrediction:PODC:2010}.

These approaches are in essence signature-based approach themselves, although they preprocess the input data in semi-supervised fashion. They may successfully increase the precision of the predictions, by training the signature-based models on pre-classified anomalies rather than plain runtime data,  but share the same limitations of
 signature-based models of requiring labeled failure data for training, and fostering predictions that overfit the failures of the training set.
Indeed, in our experiments, the \textsc{PreMiSe}\xspace approach
was ineffective against failures and failure patterns that did not correspond to the signatures considers during training. 

Guan et al.'s approach~\cite{guan:ensemble:jcom:2012} combines supervised and unsupervised learning in a different way. They exploit  Bayesian networks to automatically label the anomalous behaviors that they feed to supervised learning models for training, thus potentially relieving the burden of labelling the data.

Only few approaches work in unsupervised fashion like \textsc{Prevent}\xspace and  \textsc{EmBeD}\xspace~\cite{monni:RBM:ICST:2019} the energy-based failure prediction approach that \textsc{Prevent}\xspace exploits as state classifier and that we have already described in Section~\ref{sec:state-classifier}. 
Fernandes and al.'s~approach~\cite{Fernandes:anomalydetection:JNCA:2016} matches the actual traffic again models of the normal traffic built with either principal component analysis or the ant colony optimisation metaheuristic to detect anomalous traffic in computer networks. 
Ibidunmoye~et~al.'s~unsupervised approach~\cite{Ibidunmoye:AnomalyDetection:TNSM:2018} estimates statistical and temporal properties of KPIs during normal executions, and detects deviations by means of adaptive control charts. 
Both Bontemps~et~al.'s~\cite{Bontemps:anomalydetection:FDSE:2016} and Du~et~al.'s~\cite{Du:deeplog:CCS:2017} approaches detect anomalies with long short-term memory recurrent neural networks trained on normal execution data. 
Ahmad~et~al.'s~approach~\cite{Ahmad:HTM:Neurocomputing:2017} uses hierarchical temporal memory networks to detect anomalies in streaming data time series.

Roumani and Nwankpa~\cite{roumani:cloud-incidents:jim:2019} propose a radically different approach that extrapolates the trend of occurrence of past incidents (extracted from historical data) to predict when other incidents will likely occur again in the future, without requiring data monitored in production.

\subsection*{Failure Diagnosis}

Failure Diagnosis is the process of identifying the application components responsible for predicted failures. 
Most approaches proposed in the literature target performance bottlenecks in enterprise applications, and include
kowledge-driven, depedency-driven, observational and signature-based approaches~\cite{Ibidunmoye:AnomalyDetectionSurvey:2015}.

Knowledge-driven approaches 
rely on knowledge that analysts manually extract from historical records, and encode in rules that the inference engine processes to detect performance issues and identify the root-cause component. 
As representative example, Chung et al.'s approach~\cite{Chung:bottleneckdetection:IPDPS:2008},
is designed to work at both development- and maintenance-time, to provide testing-as-a-service, and assumes that analysts frequently update the underlying rules when observing new issues. 

Dependency-driven approaches analyze the communication flow among components, measure the frequency of the flows, and perform causal path analysis to detect anomalies and their causes. 
As a representative example, Sambasivan~et~al.'s~Spectroscope~\cite{Sambasivan:diagnosingperformance:NSDI:2011} approach assists developers to diagnose the causes of performance degrades after a software change. 
It identifies significant variations in response time or execution frequency, by comparing request flows between two corresponding executions, observed before and after the change. 

Observational approaches directly analyze the distribution of the profiled data to explain which application component correlates the most with a given system-level failure. 
As a representative example, 
Magalhaes and Silva's approach~\cite{Magalhaes:anomalydetection:NCA:2010} identifies performance anomalies of Web applications, by analyzing the correlation between the workload and response time of the transactions: A response time variation that does not correlate with a workload variation is pinpointed as a performance anomaly. 
Then, they analyze the profiled data of each component with ANOVA (analysis of variance) to spot which data (and thus which components) explain the variance in the response time~\cite{Magalhaes:rootcause:SAC:2011,Magalhaes:adaptive:NCA:2011}. 

Signature-based strategies use prior knowledge about the mapping between failures and components to diagnose failures of previously observed types~\cite{Kang:DAPA:Hot-ICE:2012}. 
As representative examples, both \textsc{PreMiSe}\xspace and \emph{ALERT} that we already mentioned above consider the failure location as an additional independent variable of their signature-based prediction models~\cite{Mariani:PreMiSE:JSS:2020,Tan:AnomalyPrediction:PODC:2010}.

All approaches suffer from restrictions that limit their applicability to distributed enterprise applications. 
Knowledge-driven and dependence-driven approaches, like Chung et al.'s approach~\cite{Chung:bottleneckdetection:IPDPS:2008} and Sambasivan et al.'s Spectroscope~\cite{Sambasivan:diagnosingperformance:NSDI:2011}, are defined to assist developers in offline analyses, for instance, after a software change.  
Signature-based approaches suffer from the same limitations of signature-based failure prediction approaches: They require labeled failure data for training, and foster predictions that overfit the failures available in the training set, without well capturing other failures and failure patterns.

Magalhaes and Silva's observational approach~\cite{Magalhaes:adaptive:NCA:2011} is the closest approach to \textsc{Prevent}\xspace anomaly ranker, since both approaches identify failing components as the components related to data strongly connected to anomalies. 
Magalhaes and Silva's approach and \textsc{Prevent}\xspace  differ in their technical core: \textsc{Prevent}\xspace uses Granger-causality centrality of the anomalous KPIs of the component, while Magalhaes and Silva rely on the ANOVA analysis. 
In this respect, \textsc{Prevent}\xspace is more failure-type-agnostic than Magalhaes and Silva's approahc, since the Granger-causality models only causal relations among KPIs, while the ANOVA analysis refer to some specific failure metric, like poor response time in Magalhaes and Silva's approach.

Overall all approaches face challenges in balancing high detection rates with low false-alarm rates, 
and the effectiveness of the different strategies largely depends on the system observability, the detection mode (real-time or post-mortem), the availability of labelled data, the dynamics of the application workload, the underlying execution context, and the nature of the collected data.

\section{Conclusions} \label{sec:conclusions}

Software failures in production are unavoidable~\cite{Gazzola:field:ISSRE:2017}. 
Predicting failures and locating failing components online are the first steps to proactively manage faults in production~\cite{Salfner:PredictSurv:ACMCompSurv:2010}.
In this paper we discuss advantages and limitations of the approaches reported in the literature, and propose \textsc{Prevent}\xspace, a novel approach that overcomes the main limitations of current approaches.

\textsc{Prevent}\xspace originally combines 
Deep autoencoder with Granger causality analysis and PageRank centrality, to effectively predict failures and locate failing components.
By relying on a combination of unsupervised approaches, \textsc{Prevent}\xspace overcomes a core limitation of supervised approaches that require seeding failures to gather labeled training data, activity hardly possible in commercial systems in production. 
  
The experimental results that we obtained on data collected by monitoring  for several weeks a popular distributed application used in commercial environments show that the original combination of unsupervised techniques in \textsc{Prevent}\xspace outperforms supervised failure prediction approaches in the majority of the experiments, and  largely reduces the unacceptable false positive rate of fault localizers used in isolation.

The main contribution of this paper are 
\begin{inparaenum}[(i)]
\item \textsc{Prevent}\xspace, an original combination of unsupervised techniques to predict failures and localize failing resources in distributed enterprise applications,  
\item a large set of data that we collected on Redis, a cluster widely used in commercial environments, data that we offer in a replication package,\footnote{The replication package at~\href{https://star.inf.usi.ch/\#/software-data/14}{https://star.inf.usi.ch/\#/software-data/14}\@\xspace}
\item a thorough evaluation that indicates the effectiveness of \textsc{Prevent}\xspace with respect to the start of the art.     
\end{inparaenum}

The results presented in this paper indicate the feasibility of unsupervised machine-learning-based approaches to predict failures and locate failing components in commercial environments, and open the horizon to study the effectiveness of unsupervised approaches for predicting failure in complex systems.   

\section*{Acknowledgement}
This work is partially supported by the Swiss SNF project ASTERIx: Automatic System TEsting of InteRactive software applications (SNF 200021\_178742), by the Italian PRIN project SISMA (PRIN 201752ENYB), and by the Italian PRIN project BigSistah (PRIN 2022EYX28N).

\begin{IEEEbiography}[{\includegraphics[width=1in,height=1.25in,clip,keepaspectratio]{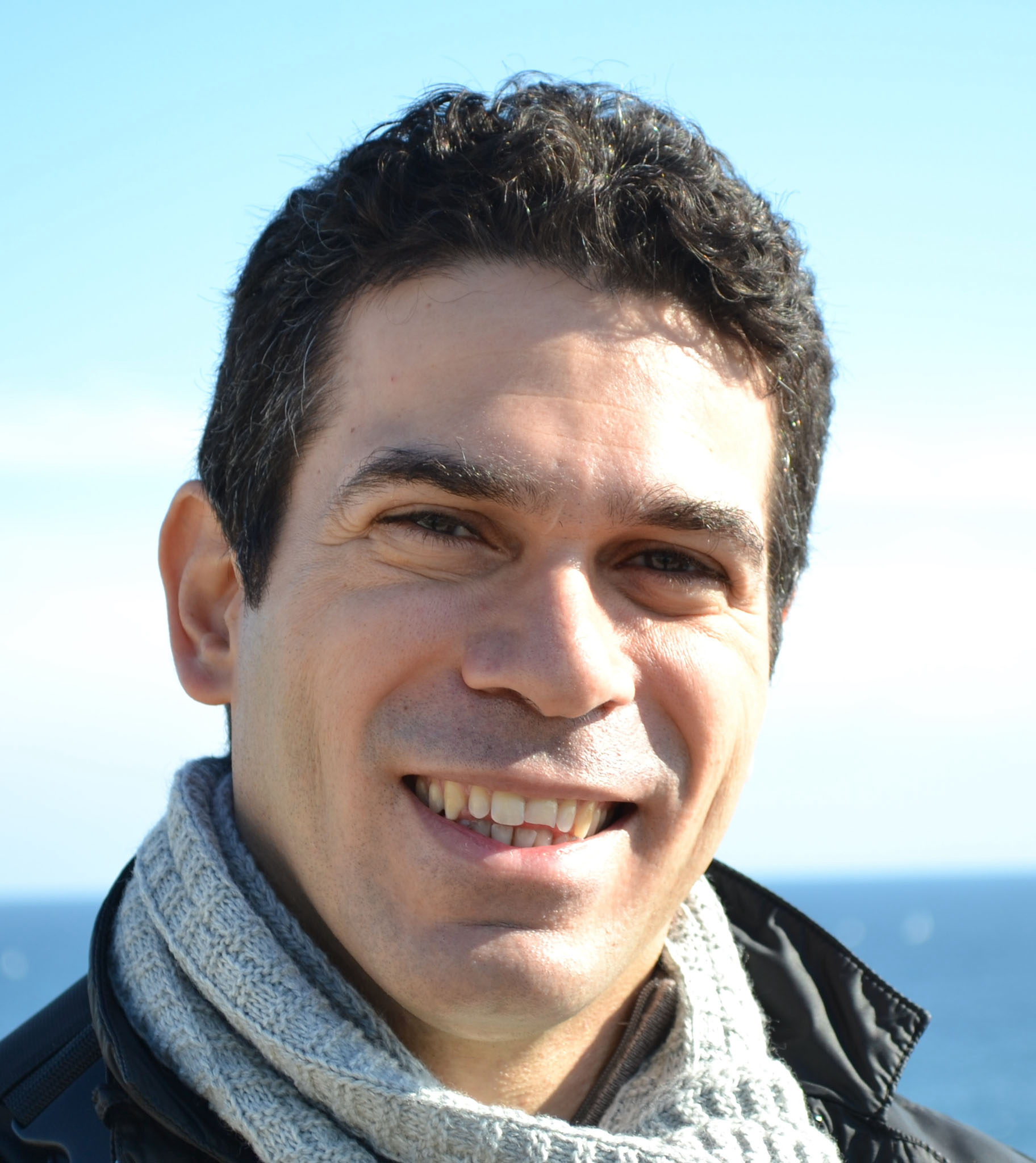}}]{Giovanni Denaro}
is Associate Professor of computer science Universit\`a degli Studi di Milano - Bicocca in Milan.
He received the Ph.D. degree in Computer Science and Engineering from Politecnico di Milano in 2002.  His research interests include software testing and analysis, formal methods for software verification and cybersecurity, distributed and service-oriented systems, and software metrics. He has been investigator in several research and development projects in  collaboration with leading European universities and companies. He is involved in the organization of major software engineering conferences.
\end{IEEEbiography}

\begin{IEEEbiography}[{\includegraphics[width=1in,height=1.25in,clip,keepaspectratio]{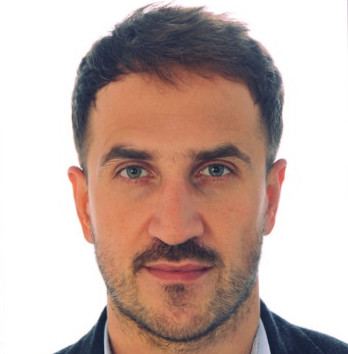}}]{Rahim Heydarov}
Rahim Heydarov is a PhD student in Computer Science at USI Universit\`a della Svizzera italiana in Lugano, Switzerland. His research is at the forefront of anomaly detection, failure prediction, and fault localization in complex software systems. His work is driven by a profound interest in harnessing the power of machine learning techniques to analyze and predict the behavior of distributed and decentralized in-cloud-deployed systems operating in non-stable environments.
\end{IEEEbiography}

\begin{IEEEbiography}[{\includegraphics[width=1in,height=1.25in,clip,keepaspectratio]{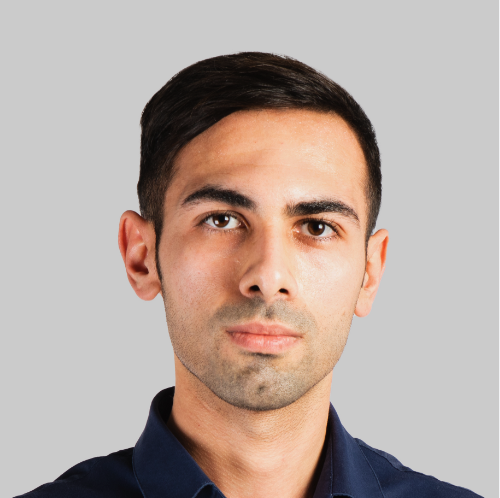}}]{Ali Mohebbi}
Ali Mohebbi received the Ph.D. in Computer Science from USI Universit\`a della Svizzera italiana, Lugano, Switzerland, in 2023. He is working on applications of Natural Language Processing and Machine Learning in software testing.  His work focuses on automatic generation of test cases for interactive applications, failure prediction and defect localization in complex cloud systems. His work is published in the proceedings  of prominent software engineering conferences, and he actively contributes as a reviewer for software engineering publications. His passion lies in bridging the gap between research and industry, striving to develop practical software engineering solutions with real-world applications.
\end{IEEEbiography}

\begin{IEEEbiography}[{\includegraphics[width=1in,height=1.25in,clip,keepaspectratio]{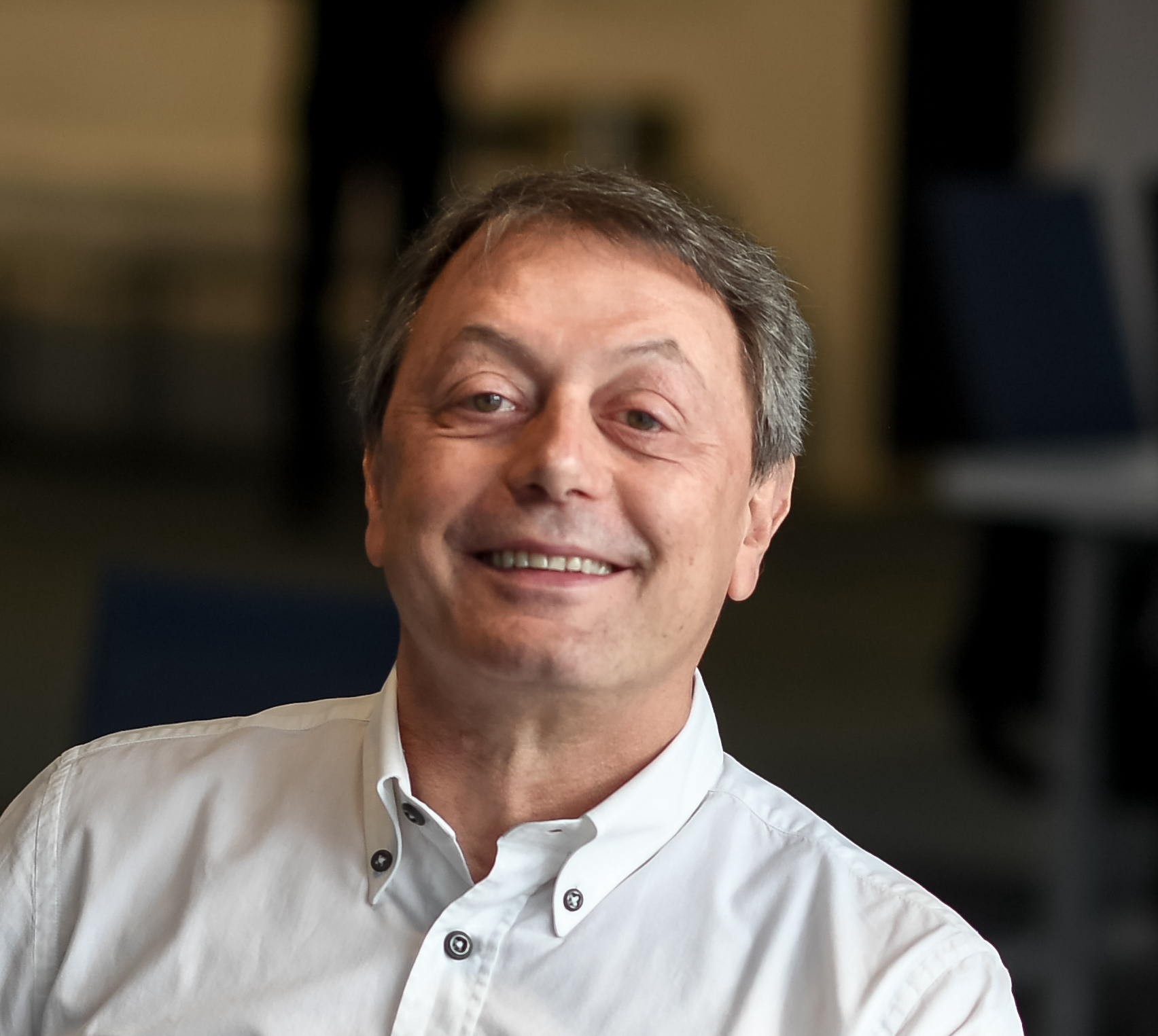}}]{Mauro Pezz\`e}
Mauro Pezz\`e is a professor of software engineering with USI Universit\`a della Svizzera Italiana in Lugano since 2006 and with Constructor Institute in Schaffhausen, since 2020.  He is professor of software engineering with Universit\`a degli Studi di Milano - Bicocca in Milan since 2000, on absence of leave since 2020. He is the editor in chief of the ACM Transactions on Software Methodologies, and has served as an associate editor of the IEEE Transactions on Software Engineering, as general chair of the ACM International Symposium on Software Testing and Analysis in 2013, program chair of the International
Conference on Software Engineering in 2012 and of the ACM International Symposium on Software Testing and Analysis in 2006. He is known for his work on software testing, program analysis, self-healing, and self-adaptive software systems. He is a senior member of the IEEE and a distinguished member of the ACM.
\end{IEEEbiography}


\begin{thebibliography}{10}

\bibitem{elasticsearch:elastic.co}
Elasticsearch reference.
\newblock \url{https://www.elastic.co/elasticsearch/}.

\bibitem{Ahmad:HTM:Neurocomputing:2017}
S.~Ahmad, A.~Lavin, S.~Purdy, and Z.~Agha.
\newblock Unsupervised real-time anomaly detection for streaming data.
\newblock {\em Neurocomputing}, 262:134--147, 2017.

\bibitem{Arnold:TCMGranger:KDD:2007}
A.~Arnold, Y.~Liu, and N.~Abe.
\newblock Temporal causal modeling with graphical granger methods.
\newblock In {\em Proceedings of the ACM SIGKDD International Conference on
  Knowledge Discovery and Data Mining}, KDD '07, pages 66--75. ACM, 2007.

\bibitem{Bach:Learning:GenerativeModelsWithoutLabeledData:ICML:2017}
S.~H. Bach, B.~He, A.~Ratner, and C.~R{\'e}.
\newblock Learning the structure of generative models without labeled data.
\newblock In {\em International Conference on Machine Learning}, pages
  273--282. PMLR, 2017.

\bibitem{Berthelot:MixmatchSemiSupervisedLearning:ANIPS:2019}
D.~Berthelot, N.~Carlini, I.~Goodfellow, N.~Papernot, A.~Oliver, and C.~A.
  Raffel.
\newblock Mixmatch: A holistic approach to semi-supervised learning.
\newblock {\em Advances in neural information processing systems}, 32, 2019.

\bibitem{Bodik:fingerprinting:EUROSYS:2010}
P.~Bodik, M.~Goldszmidt, A.~Fox, D.~B. Woodard, and H.~Andersen.
\newblock Fingerprinting the datacenter: automated classification of
  performance crises.
\newblock In {\em Proceedings of the 5th European conference on Computer
  systems}, pages 111--124, 2010.

\bibitem{bodo:process:serp:2015}
L.~Bodo, H.~Oliveira, F.~A. Breve, and D.~M. Eler.
\newblock Semi-supervised learning applied to performance indicators in
  software engineering processes.
\newblock In {\em International Conference on Software Engineering Research and
  Practice (SERP 2015), Las Vegas, EUA}, pages 255--261, 2015.

\bibitem{Bontemps:anomalydetection:FDSE:2016}
L.~Bontemps, J.~McDermott, N.-A. Le-Khac, et~al.
\newblock Collective anomaly detection based on long short-term memory
  recurrent neural networks.
\newblock In {\em International Conference on Future Data and Security
  engineering}, pages 141--152. Springer, 2016.

\bibitem{Carreira-Perpinan:ContrastiveDivergence:AISTATS:2005}
M.~{\'{A}}. Carreira{-}Perpi{\~{n}}{\'{a}}n and G.~E. Hinton.
\newblock On contrastive divergence learning.
\newblock In {\em Proc. International Workshop on Artificial Intelligence and
  Statistics}. The Society for Artificial Intelligence and Statistics, 2005.

\bibitem{Chandler:StatisticalMechanics:1987}
D.~{Chandler}.
\newblock {\em Introduction to Modern Statistical Mechanics}.
\newblock Oxford University Press, September 1987.

\bibitem{Chandola:AnomalyDetetionSurvey:CSUR:2009}
V.~Chandola, A.~Banerjee, and V.~Kumar.
\newblock Anomaly detection: A survey.
\newblock {\em ACM Computing Surveys}, 41(3):15, 2009.

\bibitem{Chapelle:SemiSupervisedLearning:MIT:2006}
O.~Chapelle, B.~Sch{\"{o}}lkopf, and A.~Zien, editors.
\newblock {\em Semi-Supervised Learning}.
\newblock The {MIT} Press, 2006.

\bibitem{Chung:bottleneckdetection:IPDPS:2008}
I.-H. Chung, G.~Cong, D.~Klepacki, S.~Sbaraglia, S.~Seelam, and H.-F. Wen.
\newblock A framework for automated performance bottleneck detection.
\newblock In {\em 2008 IEEE International Symposium on Parallel and Distributed
  Processing}, pages 1--7. IEEE, 2008.

\bibitem{colman-meixner:survey-resiliency:comst:2016}
C.~Colman-Meixner, C.~Develder, M.~Tornatore, and B.~Mukherjee.
\newblock A survey on resiliency techniques in cloud computing infrastructures
  and applications.
\newblock {\em IEEE Communications Surveys \& Tutorials}, 18(3):2244--2281,
  2016.

\bibitem{davis:failuresim:cloud:2017}
N.~A. Davis, A.~Rezgui, H.~Soliman, S.~Manzanares, and M.~Coates.
\newblock Failuresim: A system for predicting hardware failures in cloud data
  centers using neural networks.
\newblock In {\em 2017 IEEE 10th International Conference on Cloud Computing
  (CLOUD)}, pages 544--551. IEEE, 2017.

\bibitem{Dorogovtsev:CriticalPhenomenaNetworks:RevModPhys:2008}
S.~N. {Dorogovtsev}, A.~V. {Goltsev}, and J.~F.~F. {Mendes}.
\newblock Critical phenomena in complex networks.
\newblock {\em Reviews of Modern Physics}, 80:1275--1335, 2008.

\bibitem{Druck:ActiveLearning:EMNLP:2009}
G.~Druck, B.~Settles, and A.~McCallum.
\newblock Active learning by labeling features.
\newblock In {\em Proceedings of the 2009 conference on Empirical methods in
  natural language processing}, pages 81--90, 2009.

\bibitem{Du:deeplog:CCS:2017}
M.~Du, F.~Li, G.~Zheng, and V.~Srikumar.
\newblock Deeplog: Anomaly detection and diagnosis from system logs through
  deep learning.
\newblock In {\em Proceedings of the 2017 ACM SIGSAC Conference on Computer and
  Communications Security}, pages 1285--1298, 2017.

\bibitem{tehrani:threshold-sensitive:jmgs:2017}
A.~Fadaei~Tehrani and F.~Safi-Esfahani.
\newblock A threshold sensitive failure prediction method using support vector
  machine.
\newblock {\em Multiagent and Grid Systems}, 13(2):97--111, 2017.

\bibitem{Fernandes:anomalydetection:JNCA:2016}
G.~Fernandes~Jr, L.~F. Carvalho, J.~J. Rodrigues, and M.~L. Proen{\c{c}}a~Jr.
\newblock Network anomaly detection using ip flows with principal component
  analysis and ant colony optimization.
\newblock {\em Journal of Network and Computer Applications}, 64:1--11, 2016.

\bibitem{fischer:rbm-intro:springer:2012}
A.~Fischer and C.~Igel.
\newblock An introduction to restricted boltzmann machines.
\newblock In {\em Iberoamerican congress on pattern recognition}, pages 14--36.
  Springer, 2012.

\bibitem{Fulp:PredictingFailure:WASL:2008}
E.~W. Fulp, G.~A. Fink, and J.~N. Haack.
\newblock Predicting computer system failures using support vector machines.
\newblock In {\em Proceedings of the USENIX conference on Analysis of system
  logs}, WASL'08, pages 5--5. USENIX Association, 2008.

\bibitem{gao:task-failure:bigdata:2019}
J.~Gao, H.~Wang, and H.~Shen.
\newblock Task failure prediction in cloud data centers using deep learning.
\newblock {\em IEEE Transactions on Services Computing}, 2020.

\bibitem{Gazzola:field:ISSRE:2017}
L.~Gazzola, L.~Mariani, F.~Pastore, and M.~Pezz{\`e}.
\newblock An exploratory study of field failures.
\newblock In {\em Proceedings of the International Symposium on Software
  Reliability Engineering}, ISSRE~'17, 2017.

\bibitem{denaro:fault-proneness:seke:2002}
D.~Giovanni, S.~Morasca, and M.~Pezz{\`{e}}.
\newblock Deriving models of software fault-proneness.
\newblock In {\em Proceedings of the 14th International Conference on Software
  Engineering and Knowledge Engineering (SEKE 2002)}, 2002.

\bibitem{Granger:causality:Econometrica:1969}
C.~W.~J. Granger.
\newblock Investigating causal relations by econometric models and
  cross-spectral methods.
\newblock {\em Econometrica}, 37:424--438, 1969.

\bibitem{guan:ensemble:jcom:2012}
Q.~Guan, Z.~Zhang, and S.~Fu.
\newblock Ensemble of bayesian predictors and decision trees for proactive
  failure management in cloud computing systems.
\newblock {\em Journal of Communication}, 7(1):52--61, 2012.

\bibitem{Ibidunmoye:AnomalyDetectionSurvey:2015}
O.~Ibidunmoye, F.~Hern\'{a}ndez-Rodriguez, and E.~Elmroth.
\newblock Performance anomaly detection and bottleneck identification.
\newblock {\em ACM Computing Surveys}, 48(1):4:1--4:35, 2015.

\bibitem{Ibidunmoye:AnomalyDetection:TNSM:2018}
O.~Ibidunmoye, A.-R. Rezaie, and E.~Elmroth.
\newblock Adaptive anomaly detection in performance metric streams.
\newblock {\em Transactions on Network and Service Management}, 15(1):217--231,
  2018.

\bibitem{IBM:SCAPI:Tutorial:2014}
{I}{B}{M} {C}orporation.
\newblock {\em {S}mart{C}loud {A}nalytics - {P}redictive {I}nsights 1.3
  ({T}utorial)}, 2014.
\newblock Document Revision R2E2.

\bibitem{islam:failure-in-cloud:iccc:2017}
T.~Islam and D.~Manivannan.
\newblock Predicting application failure in cloud: A machine learning approach.
\newblock In {\em 2017 IEEE International Conference on Cognitive Computing
  (ICCC)}, pages 24--31. IEEE, 2017.

\bibitem{Kang:DAPA:Hot-ICE:2012}
H.~Kang, X.~Zhu, and J.~L. Wong.
\newblock Dapa: Diagnosing application performance anomalies for virtualized
  infrastructures.
\newblock In {\em 2nd $\{$USENIX$\}$ Workshop on Hot Topics in Management of
  Internet, Cloud, and Enterprise Networks and Services (Hot-ICE 12)}, 2012.

\bibitem{Laine:TemporalEnsemblingSemiSupervisedLearning:arXiv:2016}
S.~Laine and T.~Aila.
\newblock Temporal ensembling for semi-supervised learning.
\newblock {\em arXiv preprint arXiv:1610.02242}, 2016.

\bibitem{Langville:SurveyEigenvectorMethods:SIAM:2005}
A.~N. Langville and C.~D. Meyer.
\newblock A survey of eigenvector methods for web information retrieval.
\newblock {\em SIAM Review}, 47(1):135--161, 2005.

\bibitem{Liang:LearningFromMeasurements:ICML:2009}
P.~Liang, M.~I. Jordan, and D.~Klein.
\newblock Learning from measurements in exponential families.
\newblock In {\em Proceedings of the 26th annual international conference on
  machine learning}, pages 641--648, 2009.

\bibitem{lu:defect:ase:2012}
H.~Lu, B.~Cukic, and M.~Culp.
\newblock Software defect prediction using semi-supervised learning with
  dimension reduction.
\newblock In {\em Proceedings of the 27th IEEE/ACM International Conference on
  Automated Software Engineering}, pages 314--317, 2012.

\bibitem{Magalhaes:anomalydetection:NCA:2010}
J.~P. Magalhaes and L.~M. Silva.
\newblock Detection of performance anomalies in web-based applications.
\newblock In {\em 2010 Ninth IEEE International Symposium on Network Computing
  and Applications}, pages 60--67. IEEE, 2010.

\bibitem{Magalhaes:adaptive:NCA:2011}
J.~P. Magalhaes and L.~M. Silva.
\newblock Adaptive profiling for root-cause analysis of performance anomalies
  in web-based applications.
\newblock In {\em 2011 IEEE 10th International Symposium on Network Computing
  and Applications}, pages 171--178. IEEE, 2011.

\bibitem{Magalhaes:rootcause:SAC:2011}
J.~P. Magalhaes and L.~M. Silva.
\newblock Root-cause analysis of performance anomalies in web-based
  applications.
\newblock In {\em Proceedings of the 2011 ACM Symposium on Applied Computing},
  pages 209--216, 2011.

\bibitem{Malik:AutomaticDetection:ICSE:2013}
H.~Malik, H.~Hemmati, and A.~E. Hassan.
\newblock Automatic detection of performance deviations in the load testing of
  {Large} {Scale} {Systems}.
\newblock In {\em Proceedings of the International Conference on Software
  Engineering}, ICSE '13, pages 1012--1021. IEEE Computer Society, 2013.

\bibitem{Mann:SemiSupervisedLearningWeaklyLabeled:MLR:2010}
G.~S. Mann and A.~McCallum.
\newblock Generalized expectation criteria for semi-supervised learning with
  weakly labeled data.
\newblock {\em Journal of machine learning research}, 11(2), 2010.

\bibitem{Mariani:LOUD:ICST:2018}
L.~Mariani, C.~Monni, M.~Pezz{\`e}, O.~Riganelli, and R.~Xin.
\newblock Localizing faults in cloud systems.
\newblock In {\em Proceedings of the International Conference on Software
  Testing, Verification and Validation}, ICST '18, pages 262--273. IEEE
  Computer Society, 2018.

\bibitem{Mariani:PreMiSE:JSS:2020}
L.~Mariani, M.~Pezz{\`{e}}, O.~Riganelli, and R.~Xin.
\newblock Predicting failures in multi-tier distributed systems.
\newblock {\em Journal of Systems and Software}, 161, 2020.

\bibitem{Martin:Localization:PhRvE:2014}
T.~Martin, X.~Zhang, and M.~E.~J. Newman.
\newblock Localization and centrality in networks.
\newblock {\em Phys. Rev. E}, 90:052808, Nov 2014.

\bibitem{McKinney:statsmodels:Jarrodmillman:2011}
W.~McKinney, J.~Perktold, and S.~Seabold.
\newblock Time series analysis in python with statsmodels.
\newblock pages 96--102, 2011.

\bibitem{monni:energy:NIER:2019}
C.~Monni and M.~Pezz{\`{e}}.
\newblock Energy-based anomaly detection a new perspective for predicting
  software failures.
\newblock In {\em Proceedings of the 41st International Conference on Software
  Engineering: New Ideas and Emerging Results, {ICSE} {(NIER)} 2019, Montreal,
  QC, Canada, May 29-31, 2019}, pages 69--72. {IEEE} / {ACM}, 2019.

\bibitem{monni:RBM:ICST:2019}
C.~Monni, M.~Pezz{\`{e}}, and G.~Prisco.
\newblock An {RBM} anomaly detector for the cloud.
\newblock In {\em 12th {IEEE} Conference on Software Testing, Validation and
  Verification, {ICST} 2019, Xi'an, China, April 22-27, 2019}, pages 148--159.
  {IEEE}, 2019.

\bibitem{Nam:UnlabeledDefectPrediction:IEEE:2015}
J.~Nam and S.~Kim.
\newblock Clami: Defect prediction on unlabeled datasets (t).
\newblock In {\em 2015 30th IEEE/ACM International Conference on Automated
  Software Engineering (ASE)}, pages 452--463. IEEE, 2015.

\bibitem{nistor:suncat:issta:2014}
A.~Nistor and L.~Ravindranath.
\newblock Suncat: Helping developers understand and predict performance
  problems in smartphone applications.
\newblock In {\em Proceedings of the International Symposium on Software
  Testing and Analysis}, ISSTA '14, pages 282--292. ACM, 2014.

\bibitem{ozcelik:seer:tse:2016}
B.~Ozcelik and C.~Yilmaz.
\newblock Seer: {A} {Lightweight} {Online} {Failure} {Prediction} {Approach}.
\newblock {\em IEEE Transactions on Software Engineering}, 42(1):26--46, 2016.

\bibitem{Pan:TransferLearningSurvey:IEEE:2010}
S.~J. Pan and Q.~Yang.
\newblock A survey on transfer learning.
\newblock {\em IEEE Transactions on knowledge and data engineering},
  22(10):1345--1359, 2010.

\bibitem{Profillidis:ModelingTransportDemand:Elsevier:2018}
V.~A. Profillidis and G.~N. Botzoris.
\newblock {\em Modeling of transport demand: Analyzing, calculating, and
  forecasting transport demand}.
\newblock Elsevier, 2018.

\bibitem{Ratner:LargeTrainingSetsQuickly:ANIPS:2016}
A.~J. Ratner, C.~M. De~Sa, S.~Wu, D.~Selsam, and C.~R{\'e}.
\newblock Data programming: Creating large training sets, quickly.
\newblock {\em Advances in neural information processing systems}, 29, 2016.

\bibitem{roumani:cloud-incidents:jim:2019}
Y.~Roumani and J.~K. Nwankpa.
\newblock An empirical study on predicting cloud incidents.
\newblock {\em International journal of information management}, 47:131--139,
  2019.

\bibitem{Salfner:PredictSurv:ACMCompSurv:2010}
F.~Salfner, M.~Lenk, and M.~Malek.
\newblock A survey of online failure prediction methods.
\newblock {\em ACM Computing Surveys}, 42(3):1--42, 2010.

\bibitem{Sambasivan:diagnosingperformance:NSDI:2011}
R.~R. Sambasivan, A.~X. Zheng, M.~De~Rosa, E.~Krevat, S.~Whitman, M.~Stroucken,
  W.~Wang, L.~Xu, and G.~R. Ganger.
\newblock Diagnosing performance changes by comparing request flows.
\newblock In {\em NSDI}, volume~5, pages 1--1, 2011.

\bibitem{Sauvanaud:FaultLocalizationClearwater:ISSRE:2016}
C.~Sauvanaud, K.~Lazri, M.~Kaâniche, and K.~Kanoun.
\newblock Anomaly detection and root cause localization in virtual network
  functions.
\newblock In {\em Proceedings of the International Symposium on Software
  Reliability Engineering}, ISSRE '16, pages 196--206. IEEE Computer Society,
  2016.

\bibitem{Scott:SNAnalysis:book:2011}
J.~P. Scott and P.~J. Carrington.
\newblock {\em The SAGE Handbook of Social Network Analysis}.
\newblock Sage Publications Ltd., 2011.

\bibitem{Seabold:statsmodels:SciPy:2010}
S.~Seabold and J.~Perktold.
\newblock Statsmodels: Econometric and statistical modeling with python.
\newblock In {\em Proceedings of the 9th Python in Science Conference},
  volume~57, page~61. Austin, TX, 2010.

\bibitem{stack:selfhealing:CloudNG:2017}
P.~Stack, H.~Xiong, D.~Mersel, M.~Makhloufi, G.~Terpend, and D.~Dong.
\newblock Self-healing in a decentralised cloud management system.
\newblock In {\em Proceedings of the 1st International Workshop on Next
  generation of Cloud Architectures, CloudNG@EuroSys 2017, Belgrade, Serbia,
  April 23-26, 2017}, pages 3:1--3:6, 2017.

\bibitem{Stewart:LabelFreeSupervision:AAAI:2017}
R.~Stewart and S.~Ermon.
\newblock Label-free supervision of neural networks with physics and domain
  knowledge.
\newblock In {\em Thirty-First AAAI Conference on Artificial Intelligence},
  2017.

\bibitem{sun:hardware-failure:dac:2019}
X.~Sun, K.~Chakrabarty, R.~Huang, Y.~Chen, B.~Zhao, H.~Cao, Y.~Han, X.~Liang,
  and L.~Jiang.
\newblock System-level hardware failure prediction using deep learning.
\newblock In {\em 2019 56th ACM/IEEE design automation conference (DAC)}, pages
  1--6. IEEE, 2019.

\bibitem{Tan:AnomalyPrediction:PODC:2010}
Y.~Tan, X.~Gu, and H.~Wang.
\newblock Adaptive system anomaly prediction for large-scale hosting
  infrastructures.
\newblock In {\em Proceedings of the Symposium on Principles of Distributed
  Computing}, PODC '12, pages 173--182. ACM, 2010.

\bibitem{tan:prepare:icdcs:2012}
Y.~Tan, H.~Nguyen, Z.~Shen, X.~Gu, C.~Venkatramani, and D.~Rajan.
\newblock Prepare: Predictive performance anomaly prevention for virtualized
  cloud systems.
\newblock In {\em 2012 IEEE 32nd International Conference on Distributed
  Computing Systems}, pages 285--294. IEEE, 2012.

\bibitem{tu:frugal:ase:2021}
H.~Tu and T.~Menzies.
\newblock {FRUGAL}: unlocking semi-supervised learning for software analytics.
\newblock In {\em 2021 36th IEEE/ACM International Conference on Automated
  Software Engineering (ASE)}, pages 394--406. IEEE, 2021.

\bibitem{Zhang:WeakSupervision:CoRR2022}
J.~Zhang, C.~Hsieh, Y.~Yu, C.~Zhang, and A.~Ratner.
\newblock A survey on programmatic weak supervision.
\newblock {\em CoRR}, abs/2202.05433, 2022.

\bibitem{zhang:lowrank:iet:2021}
Z.-W. Zhang, X.-Y. Jing, and F.~Wu.
\newblock Low-rank representation for semi-supervised software defect
  prediction.
\newblock {\em IET Software}, 12(6):527--535, 2018.

\end{thebibliography}
\end{document}